\documentclass[11pt,a4paper]{article}

\usepackage[utf8]{inputenc}
\usepackage[T1]{fontenc}
\usepackage[english]{babel}

\usepackage{indentfirst}
\usepackage{amsmath,mathtools}
\usepackage{amssymb}
\usepackage{graphicx}
\usepackage{subfigure}
\usepackage{psfrag}
\usepackage{color}

\allowhyphens
\newcommand\p            {\partial}
\newcommand\Psid         {\Psi^{\dagger}}

\renewcommand\vec[1]{{\vert{#1}\rangle}}
\newcommand\cev[1]{{\langle{#1}\vert}}
\newcommand\vac{{\vec 0}}
\newcommand\cav{{\cev 0}}

\newcommand{\valos}{\mathbb{R}}

\newcommand{\eps}{\varepsilon}

\newcommand{\bizveg}{\hfill $\blacksquare$}

\newcommand{\ordo}{\mathcal{O}}

\newtheorem{thm}{Theorem}

\newcommand{\ket}[1]{{\left|#1\right\rangle}}
\newcommand{\bra}[1]{{\left\langle #1\right|}}

\newcommand{\skalarszorzat}[2]{{\langle #1 | #2 \rangle}}

\newcommand{\vev}[1]{\left\langle #1 \right\rangle}

\newcommand{\ABAB}{\mathbb{B}}
\newcommand{\ABAC}{\mathbb{C}}

\usepackage{ifpdf}

\ifpdf
\usepackage{epstopdf}
\usepackage[pdftex,ps2pdf,dvips,colorlinks,urlcolor=blue,citecolor=blue,linkcolor=blue]{hyperref}
\else
\usepackage[hypertex,ps2pdf,dvips,colorlinks,urlcolor=blue,citecolor=blue,linkcolor=blue]{hyperref}
\fi

\makeatother
\numberwithin{equation}{section}
\numberwithin{thm}{section}

\title{
Mean values of local operators in highly excited Bethe states
}
\author{Bal\'azs Pozsgay\\
Institute for Theoretical Physics, Universiteit van Amsterdam\\
Science Park 904, Postbus 94485, 1090 GL Amsterdam, The Netherlands}
\date{September 2010}

\begin{document}

\maketitle

\begin{abstract}
We consider expectation values of local operators in (continuum)
integrable models in a situation when the mean value is
calculated in a single Bethe state with a large number of particles. We
develop a form factor expansion for the thermodynamic limit of the
mean value, which applies whenever the distribution of Bethe roots is given
by smooth density functions. We present three applications of our
general result: i)  In the framework of integrable Quantum Field
Theory (IQFT)
we present a derivation of the LeClair-Mussardo formula for
finite temperature one-point functions. We also extend the results to
boundary operators in Boundary Field Theories.
ii) We establish the
LeClair-Mussardo formula for the non-relativistic 1D Bose gas in the
framework of Algebraic Bethe Ansatz (ABA). This way we obtain an alternative
derivation of the results of Kormos et. al. for the (temperature dependent)
local correlations using only the concepts of ABA.  iii)
In IQFT we consider 
the long-time limit
of  one-point functions after a certain type of global quench. It is
shown that our general results imply the integral series found by
Fioretti and Mussardo.
We also discuss 
the generalized Eigenstate Thermalization Hypothesis in the context of
quantum quenches in integrable models. It is shown that a single mean
value always takes the form of a thermodynamic average in a Generalized
Gibbs Ensemble, although the relation to the conserved charges is
rather indirect.
\end{abstract}

\section{Introduction and our main result}

One of the central tasks of many-body quantum physics is the calculation of correlation functions.
In addition to the ground state properties it is also important to consider correlations calculated
in excited states. These quantities are relevant in (at least)
two physical situations: they describe the finite temperature properties
of the system and they can be used to calculate observables after a sudden quantum
quench \cite{quench-reviews}. Besides calculating the space and time dependent correlations
it is also important to consider the one-point functions, 
ie. the mean values of local operators. Important examples include the
(properly regularized) products of field operators, which can describe
densities of conserved charges or certain amplitudes of inelastic
many-body processes \cite{g3,low-D-trapped}. 

In this paper we consider one-point functions in one-dimensional
integrable models. These theories can be solved by
the Bethe Ansatz, which means that the spectrum and also the thermodynamic
quantities can be determined exactly. Moreover, there are powerful
methods available to obtain correlation
functions. One of the most general methods is the so-called form factor approach, which
consists of two main steps. 
The first step is the calculation of the matrix elements (form
factors) of the local operators in the eigenstate basis of the
Hamiltonian, which can then be used to obtain the correlation function
as the sum of a spectral series.

In this work we elaborate on the form factor expansion for one-point
functions, more specifically  we consider mean values in highly excited states with a
smooth distribution of Bethe roots. We only consider theories with purely
diagonal scattering, but otherwise the calculations are quite
general. We consider both relativistic Quantum Field
Theories (QFT's)  and non-relativistic many-body systems in the framework of
Algebraic Bethe Ansatz. There are three main sources of motivation for
the present work.

\medskip

The first motivation comes from integrable QFT, where the form factors
are obtained as solutions to the so-called form factor bootstrap
program
\cite{Karowski:1978vz,smirnov_ff,zam_Lee_Yang,Delfino:1996nf}. Originally
the form factor approach was developed to obtain 
two-point functions at zero temperature, but
it was hoped for that the form
factors could be used to describe finite temperature correlations
too
\cite{Leclair:1996bf,leclair_mussardo,Saleur:1999hq,Delfino:2001sz,Mussardo:2001iv}. The
formal Boltzmann averages  
are plagued with divergences, which arise from
singularities of the form factors and different
contributions to the partition function. It is fairly easy to show
that the divergences cancel, but the determination of the left-over
pieces is in general a demanding problem. 

In \cite{leclair_mussardo} LeClair and Mussardo proposed compact integral formulas for
the finite temperature correlations, which
involve the (appropriately regularized) zero-temperature form factors
and the occupation numbers as determined from the Thermodynamic Bethe
Ansatz (TBA). In the case of two-point functions the results of \cite{leclair_mussardo} 
 were questioned in
\cite{Saleur:1999hq,CastroAlvaredo:2002ud}. On the other hand, it was
proven it \cite{Saleur:1999hq} that for one-point functions the
LeClair-Mussardo (LM) formula is correct for operators describing densities
of conserved quantities. Moreover, it was shown in \cite{fftcsa2}
using a finite-volume regularization that
the LM series is valid for arbitrary operators up to the first three
non-trivial orders. 
However, the methods of \cite{fftcsa2} can only be applied order-by-order, and
the calculations become rather tedious for higher orders, therefore they
are not suitable for an all-order-proof of the LM formula.

\medskip

The second motivation for the present work comes from recently
discovered relations between integrable QFT and non-relativistic
models of Algebraic Bethe Ansatz (ABA). In \cite{sinhG-LL1,sinhG-LL2}
a non-relativistic limit of the sinh-Gordon field theory was applied
to obtain physical observables in the Lieb-Liniger model
\cite{Lieb-Liniger,korepinBook}. They performed the non-relativistic
limit of the LM series 
for certain sinh-Gordon operators to arrive at the (temperature
dependent) local correlations in the Bose gas. It was pointed out
later in \cite{nonrelFF} that the non-relativistic limit not only
applies to the physical observables, but also to the individual form
factors entering the integral series. This means that the final
results of \cite{sinhG-LL1,sinhG-LL2} can be formulated using only
the quantities calculated in the Bose gas, without actually referring to the
sinh-Gordon model. Evidently, this suggests that the LeClair-Mussardo
formalism is not at all limited to relativistic field theory and it
could be applicable to other Bethe Ansatz solvable models too. We
stress that although there is a vast literature devoted to 
correlation functions in the Bose gas, the papers \cite{sinhG-LL1,sinhG-LL2}
 were apparently the first ones to calculate
the many-body local correlations using Bethe Ansatz
techniques.

It was a vital point of \cite{sinhG-LL1,sinhG-LL2} to apply the
LeClair-Mussardo formalism in the case of an arbitrary prescribed
particle density, or in other words in the presence of a non-zero chemical
potential $\mu$. Although this seems to be quite natural, we
remark that neither the original derivation in
\cite{leclair_mussardo}, nor the partial proof of \cite{fftcsa2}
applies in this case. Indeed, the calculation of \cite{fftcsa2}
heavily relied on the fact that if $\mu=0$, then the particle density $\rho=N/L$
can be expanded into a low-temperature expansion and that
$\rho(T=0)=0$. On the other hand, the success of
\cite{sinhG-LL1,sinhG-LL2} 
suggests that it should be possible to prove the LM
series using a different approach, 
which should work at arbitrary $\mu$, and both in relativistic QFT and
in Algebraic Bethe Ansatz.

\bigskip

The third motivation for the present work is the quench problem of
Fioretto and Mussardo \cite{davide}, where the authors consider 
the long-time limit of one-point functions after
a sudden global quench. The resulting integral series is a
generalization of the LM formula, where
the statistical weight functions do not follow from thermodynamic
quantities, but from the microscopic
amplitudes of the quantum quench and a generalized TBA-like dressing
procedure. Once again, the compact form of the final result 
suggests that it should be possible to derive it using alternative
methods, bypassing the combinatorial difficulties 
encountered in \cite{davide}.

\bigskip

In this paper we approach the above mentioned problems using a finite
volume regularization for the mean values. 
We start from the Boltzmann average for thermal one-point functions in
a finite volume $L$, which can be
written as
\begin{equation}
  \label{eq:Gibbs}
  \vev{\ordo}_{T}=\frac{\sum_i \bra{i}\ordo\ket{i}_Le^{-E_i/T}}{\sum_i
    e^{-E_i/T}}
\end{equation}
where the summation runs over a complete set of states. Note that
contrary to the infinite volume case the above formula is perfectly
well-defined in finite volume. For simplicity we assume that there is
only one particle type in the spectrum; the generalization to
arbitrary diagonal scattering theories is straightforward. In this
case the Bethe equations (the quantization conditions for the finite
volume states $\ket{i}_L$ ) can be written as
\begin{equation*}
  e^{ip(\theta_j)L}\prod_{k\ne j}S(\theta_j-\theta_k)=1,\qquad
  j=1\dots N
\end{equation*}
Here $S(\theta)$ is the elastic S-matrix which is a pure phase in this
case. The variables $\theta_j$ are the rapidity
parameters and one-particle momenta and energies are given by the
functions $p(\theta)$ and $e(\theta)$. The results to be presented
below apply both to relativistic and non-relativistic situations.

In \cite{fftcsa2} a low-temperature expansion of
\eqref{eq:Gibbs} was performed in massive integrable field theory. The
derivation was built on the fact that if the chemical
potential is zero then the contribution of the $N$-particle sector of
the Fock space can be estimated as $(mLe^{-m/T})^N$, where $L$ is the
volume and $m$ is the mass of the particle. Therefore, in the regime
\begin{equation}
\label{regime1}
  1\ll mL \ll e^{m/T}
\end{equation}
the volume can be large enough to replace sums with appropriate
integrals, and in the same time
the relevant contributions in \eqref{eq:Gibbs} would still come from sectors with
a low number of particles. This was used in \cite{fftcsa2} to derive
a systematic low-temperature expansion and the $L\to\infty$ limit was taken at
the end of the calculations. The same ideas were employed in the
papers \cite{Konik:2001gf,Essler:2007jp,Essler:2009zz,D22}.

In the present work we pursue a
complementary approach: we consider a very large volume such that
there are a large number of particles present in the system. 
In this case it is expected that 
the thermal average will be dominated by those states in which the
distribution of Bethe roots is described by the infinite volume TBA
equations \cite{YangYang1,YangYang2,zam-tba}. In fact the TBA yields densities which minimize
the free energy functional for the partition function, and one can
argue that the insertion of the local operator does not shift this
saddle point solution. Moreover, one can show that in the
thermodynamic limit it is sufficient to 
consider only one particular state and evaluate the mean value on
this state.

Let us therefore introduce a ``thermal state'' 
as
\begin{equation*}
 \ket{\Omega}_L=\ket{\theta_1,\dots,\theta_N}_L\qquad\qquad
_L\skalarszorzat{\Omega}{\Omega}_L=1,
\end{equation*}
where it is understood that the rapidities $\theta_j$ satisfy the BA equations, the
particle number is  $N=[\rho L]$ with $\rho$ being the infinite
volume particle density ($[x]$ denotes the integer part of $x$), and the distribution of roots is 
given by the TBA equations (detailed 
definitions will be given below). Then the mean value \eqref{eq:Gibbs} is given 
by the thermodynamic limit 
\begin{equation}
\label{diag-ff}
  \vev{\ordo}_{T}=\lim_{N,L\to\infty} \bra{\Omega}\ordo\ket{\Omega}_L
\end{equation}
In general the matrix elements with a large number of particles are
complicated objects, and the evaluation of the thermodynamic limit can be
a difficult problem \cite{springerlink:10.1007/BF01029221,2009JMP....50i5209K,2010arXiv1003.4557K}.
However, in the case of mean values the situation is much simpler.
In \cite{fftcsa2} it was shown that a diagonal matrix element is the sum
of its properly defined disconnected terms, which can be expressed 
using form factors with a lower number
of particles. Then the remaining task is to perform the thermodynamic
limit for the disconnected terms.

It is useful to investigate the mean value in a more general
situation. 
Let us consider a Bethe state $\ket{\psi}_L$ in a large volume with
$N=[\rho L]$ particles.
We assume that the distribution of
Bethe roots is given by the functions $\rho^{(o)}(\theta)$  and
$\rho^{(h)}(\theta)$, which describe the densities of occupied states
and holes, respectively. They satisfy 
the constraint
\begin{equation}
\label{alap-density}
    \rho^{(o)}(\theta)+ \rho^{(h)}(\theta)=\frac{1}{2\pi} p'(\theta)
   +\int_{-\infty}^\infty\frac{d\theta'}{2\pi} 
\varphi(\theta-\theta')\rho^{(o)}(\theta')
\end{equation}
where $\varphi=-i \frac{d}{d\theta} \log S(\theta)$ is the scattering
kernel and $p'(\theta)$ is the derivative of the one-particle momentum.  The particle
density is then given by the integral
\begin{equation*}
  \rho=\int {d\theta}\ \rho^{(o)}(\theta)
\end{equation*}
We consider the thermodynamic limit of the mean value:
\begin{equation}
\label{TwinPeaks}
   \vev{\ordo}=\lim_{N,L\to\infty} \bra{\psi}\ordo\ket{\psi}_L
\end{equation}
For the above limit we find our main result:
\begin{equation}
  \label{Apollo11}
  \begin{split}
 \vev{\ordo}=\sum_n \frac{1}{n!}
\int_{-\infty}^\infty \frac{d\theta_1}{2\pi}\dots  \frac{d\theta_n}{2\pi}
\left(\prod_{j=1}^n f(\theta_j)\right)
F^\ordo_{2n,c}(\theta_1,\dots,\theta_n)
\end{split}
\end{equation}
Here $F^\ordo_{2n,c}$ are regularized evaluations of the diagonal
form factors (to be defined in the main text) and the statistical weight functions are given by 
\begin{equation}
  \label{fdef}
f(\theta)=\frac{\rho^{(o)}(\theta)}{\rho^{(h)}(\theta)+\rho^{(o)}(\theta)}.
\end{equation}
The $n$th term in the integral series arises from the $n$-particle
disconnected terms of the original matrix element
\eqref{TwinPeaks}. The form factors describe $n$-particle
processes over the Fock vacuum and the only effect of the non-trivial
background is the appearance of the weight functions $f(\theta)$.
Remarkably \eqref{Apollo11} holds both in
massive relativistic Field Theories and in the non-relativistic 1D
Bose gas, which is a gapless system. This will be proven in sections
\ref{sec:FF} and \ref{LLsec}, respectively.
In the latter case
\eqref{Apollo11} can be considered as an (improved) $1/\gamma$
expansion, where $\gamma$ is the dimensionless effective coupling
of the Bose gas.

In the finite temperature problem one introduces the
pseudo-energy function $\eps(\theta)$ as
\begin{equation*}
  \frac{\rho^{(o)}(\theta)}{\rho^{(h)}(\theta)}=e^{-\eps(\theta)},
\end{equation*}
where $\eps(\theta)$ is determined by the TBA equation
\begin{equation}
\label{TBA}
  T \eps(\theta)=e(\theta)-\mu-T
  \int_{-\infty}^\infty\frac{d\theta'}{2\pi} 
\varphi(\theta-\theta') \log(1+e^{-\eps(\theta')}),
\end{equation}
with $e(\theta)$ being the one-particle energy. Applying
\eqref{Apollo11} in this case one obtains
\begin{equation}
\label{LM}
    \vev{\ordo}_{T,\mu}=\sum_m \frac{1}{n!}
\int \frac{d\theta_1}{2\pi}\dots  \frac{d\theta_n}{2\pi}
\left(\prod_j \frac{1}{1+e^{\eps(\theta_j)}}\right)
F^\ordo_{2n,c}(\theta_1,\dots,\theta_n),
\end{equation}
which is just the LeClair-Mussardo series as originally proposed in
\cite{leclair_mussardo}. Note that the general result \eqref{Apollo11}
allows for an arbitrary chemical potential $\mu$. As a special case one
can take the zero temperature limit with a fixed particle density to
obtain
\begin{equation}
\label{LM-T0}
    \vev{\ordo}_{T=0,\mu}=\sum_k \frac{1}{n!}
\int_{-B}^B \frac{d\theta_1}{2\pi}\dots \int_{-B}^B  \frac{d\theta_n}{2\pi}
F^\ordo_{2n,c}(\theta_1,\dots,\theta_n) ,
\end{equation}
where $B$ is the Fermi-rapidity. This remarkably simple and intuitive
result was written down for the first time in
\cite{sinhG-LL1,sinhG-LL2}; it applies whenever the ground state with
a fixed number of particles is unique. 

Let us consider the (typically infinite) set of local onserved quantities
$Q_i$ with $i=1,2,\dots,\infty$. The first two members of the series can be
chosen as the particle number and the energy.
If the one-particle eigenvalues are given by
the functions $q_i(\theta)$ then the macroscopic values of the
 charges are given by the integrals
\begin{equation}
\label{charges}
   Q_i=L \int d\theta \rho^{(o)}(\theta)\ q_i(\theta)
\end{equation}
Assuming that there is a one-to-one correspondence between the root
densities and the conserved charges (ie. the relation
\eqref{charges} can be inverted) the main result \eqref{Apollo11} can be interpreted as
a ``Generalized Eigenstate Thermalization Hypothesis''  \cite{2008Natur.452..854R}: the mean value
always takes the form of a thermodynamic average and the weight functions
only depend on the macroscopic value of the conserved charges. 

The remainder of the paper will be devoted to the proofs and
applications of our general formula \eqref{Apollo11}. In section \ref{sec:FF} we
establish \eqref{Apollo11} by performing the thermodynamic limit of
the mean value in relativistic
QFT.
In \ref{sec:boundary} we extend the results to boundary operators in
Boundary QFT and prove the corresponding LeClair-Mussardo formula
proposed in \cite{Takacs:2008ec}.
In section \ref{LLsec} we apply our formalism to the
Lieb-Liniger model in the framework of Algebraic Bethe Ansatz. 
In \ref{sec:quench} we explain how to apply
our general formula to quench problems. We re-derive the results of Fioretto and Mussardo
\cite{davide}  and we also discuss more general quench situations.
Section \ref{sec:conclusions} includes our conclusions, and a number of
technical details are presented in the appendices.

\section{Form factors and expectation values}

\label{sec:FF}

Let us consider a massive integrable QFT.
The basic object of the theory is the factorized $S$-matrix \cite{zam-zam,Mussardo:1992uc}. 
 For simplicity
we assume that there is only one particle type in the spectrum with no
internal degrees of freedom. 
In this case the $S$-matrix is a pure phase:
\begin{equation*}
  S(\theta_i-\theta_j)=e^{i\vartheta(\theta_i-\theta_j)}
\end{equation*}
One-particle energy and momentum are given by the functions
\begin{equation*}
  e(\theta)=m\cosh\theta\qquad\text{and}\qquad
p(\theta)=m\sinh\theta
\end{equation*}
In infinite volume the Hilbert space is spanned by the asymptotic
states
\begin{equation*}
  \ket{\theta_1,\dots,\theta_N}
\end{equation*}
Their energy and momentum (and higher conserved charges) can be
calculated additively. 

Consider a local operator $\ordo(x,t)$. The form factors of $\ordo$
are defined as \cite{smirnov_ff}
\begin{equation}
\label{FFdef}
F^\ordo_{M,N}(\theta'_1,\dots,\theta'_M|\theta_1,\dots,\theta_N)=
\bra{\theta'_1,\dots,\theta'_M}\ordo\ket{\theta_1,\dots,\theta_N}
\end{equation}
With the help of the crossing relations
\begin{eqnarray}
 &  &
 F_{M,N}^{\mathcal{O}}(\theta_{1}^{'},\dots,\theta_{M}^{'}|\theta_{1},\dots,\theta_{N})= 
F_{M-1,N+1}^{\mathcal{O}}(\theta_{1}^{'},\dots,\theta_{M-1}^{'}|\theta_{M}^{'}+i\pi,\theta_{1},\dots,\theta_{N})\\
 &  &
 +\sum_{k=1}^{N}2\pi\delta(\theta_{M}^{'}-\theta_{k})\prod_{l=1}^{k-1}S(\theta_{l}-\theta_{k}) 
F_{M-1,N-1}^{\mathcal{O}}(\theta_{1}^{'},\dots,\theta_{M-1}^{'}
|\theta_{1},\dots,\theta_{k-1},\theta_{k+1}\dots,\theta_{N})\nonumber 
\label{eq:ffcrossing}
\end{eqnarray}
all form factors can be expressed in terms of the elementary form
factors
\begin{equation}
F_{N}^{\mathcal{O}}(\theta_{1},\dots,\theta_{N})=
\langle0\vert\mathcal{O}(0,0)\vert\theta_{1},\dots,\theta_{N}\rangle
\end{equation}
which satisfy the following equations: 

I. Lorentz transformation: 

\begin{equation}
F_{N}^{\mathcal{O}}(\theta_{1}+\Lambda,\theta_{2}+\Lambda,\dots,\theta_{N}+\Lambda)
=\exp\left(s_\mathcal{O}\Lambda\right)F_{N}^{\mathcal{O}}(\theta_1,\theta_{2},\dots,\theta_{N})\label{eq:shiftaxiom}\end{equation}
where $s_\mathcal{O}$ denotes the Lorentz spin of the operator $\mathcal{O}$.

II. Exchange:

\begin{eqnarray}
 &  & F_{N}^{\mathcal{O}}(\theta_{1},\dots,\theta_{k},\theta_{k+1},\dots,\theta_{N})=\nonumber \\
 &  & \qquad S(\theta_{k}-\theta_{k+1})F_{N}^{\mathcal{O}}(\theta_{1},\dots,\theta_{k+1},\theta_{k},\dots,\theta_{N})
\label{eq:exchangeaxiom}
\end{eqnarray}

III. Cyclic permutation: \begin{equation}
F_{N}^{\mathcal{O}}(\theta_{1}+2i\pi,\theta_{2},\dots,\theta_{N})
=F_{N}^{\mathcal{O}}(\theta_{2},\dots,\theta_{N},\theta_{1})\label{eq:cyclicaxiom}\end{equation}

IV. Kinematical singularity\begin{equation}
-i\mathop{\textrm{Res}}_{\theta=\theta^{'}}
F_{N+2}^{\mathcal{O}}(\theta+i\pi,\theta^{'},\theta_{1},\dots,\theta_{N})
=\left(1-\prod_{k=1}^{N}S(\theta'-\theta_{k})\right)F_{N}^{\mathcal{O}}(\theta_{1},\dots,\theta_{N})
\label{eq:kinematicalaxiom}\end{equation}

There is also a further equation related to bound states which we do not
need in the sequel. 

\bigskip

In order to derive a regularized form factor expansion for the mean
values we consider the theory in a finite volume $L$. 
Assuming periodic
boundary conditions the Bethe equations are written as
\begin{equation*}
  e^{ip_jL}\prod_{k\ne j}S(\theta_j-\theta_k)=1,\qquad
  j=1\dots N,
\end{equation*}
where $p_j=p(\theta_j)$. In the logarithmic form:
\begin{equation}
\label{logBY}
  Q_j=p_jL+\sum_{k\ne j}\vartheta(\theta_j-\theta_k)=2\pi
  I_j\qquad j=1\dots N,\quad I_j\in \mathbb{Z}
\end{equation}
The multi-particle energies and momenta are calculated additively:
\begin{equation*}
  P_N=\sum_j p(\theta_j)\qquad
E_N=\sum_j e(\theta_j)
\end{equation*}
In the following we denote the finite volume states as
\begin{equation*}
  \ket{\theta_1,\dots,\theta_N}_L
\end{equation*}
where it is understood that the rapidities solve eq. \eqref{logBY}.

It is a very important task to determine the finite volume form
factors
in terms of the infinite volume quantities. In
the case of relativistic QFT this problem was solved in
\cite{fftcsa1,fftcsa2}. In a generic case the result reads
\begin{equation}
   \bra{\theta'_1,\dots,\theta'_M}\ordo
   \ket{\theta_1,\dots,\theta_N}_L=
\frac{F^\ordo_{N,M}(\theta'_1,\dots,\theta'_M|\theta_1,\dots,\theta_N)}
{\sqrt{\rho_M(\theta'_1,\dots,\theta'_M)\rho_N(\theta_1,\dots,\theta_N)}}
+\ordo(e^{-\mu L})
\label{FVFF}
\end{equation}
The quantities $\rho_{N}$ and $\rho_M$ are the multi-particle density of states
and they are given by
\begin{equation}
  \rho_N(\theta_1,\dots,\theta_N)=\det \mathcal{J}^{ij},
\qquad\qquad
\mathcal{J}^{ij}=\frac{\partial Q_i}{\partial \theta_j}
\label{Jdef}
\end{equation}
The error exponent $\mu$ is universal in the sense that it only
depends on the analytic properties of the S-matrix and not on the
particular form factor in question. 
Equation \eqref{FVFF} is very
natural in the Algebraic Bethe Ansatz approach for 
 models with $sl(2)$ or $U_q(sl(2))$ symmetry
\cite{korepin-slavnov}. In these models \eqref{FVFF}
is exact and $\rho_N$ is just the
Gaudin-determinant, which describes the norm of the Bethe wave
function.

The relation \eqref{FVFF} is valid when there are no kinematical poles
present due to colliding rapidities.
It was pointed out in \cite{fftcsa2}
that there are only two situations when rapidities do coincide in a finite
volume. One possibility is to have
zero-momentum particles in parity symmetric states; these matrix
elements will not be needed here, therefore we omit the details which
can be found in \cite{fftcsa2}. The other case is the problem of
diagonal matrix elements (expectation values),
which will be discussed below. Note that the
diagonal limit in \eqref{FVFF} is ill-defined due to the multiple
kinematical poles prescribed by \eqref{eq:kinematicalaxiom}. 

In the following
we restate the relevant results of \cite{fftcsa2}.
Consider the diagonal form factor of $N$ particles in infinite volume:
\begin{equation}
\label{altalanosdiag}
  F_{2N}^\ordo(\theta_1+\eps_1,\dots,\theta_N+\eps_N|\theta_N,\dots,\theta_1)
\end{equation}
Here the singularities have been shifted off by the infinitesimal
quantities $\eps_i$. It was proven in \cite{fftcsa2} that there exists a finite limit
when all $\eps_i$ go to zero simultaneously with their ratios
fixed. Moreover, there are two special evaluation schemes which respect the
physical requirement that the diagonal form factors should not depend on the order of the
rapidities. 
First of all,
one can consider the symmetric limit
\begin{equation}
\label{symmetric-def}
F_{2N,s}^\ordo(\theta_1,\dots,\theta_N)\equiv
\lim_{\eps\to 0}  F_{2N}^\ordo(\theta_1+\eps,\dots,\theta_N+\eps|\theta_N,\dots,\theta_1)
\end{equation}
On the other hand,  one can also consider
the connected part of the diagonal form factor which is defined to be the
contribution to  \eqref{altalanosdiag} which does not contain any
singular factors of the form $\eps_i/\eps_j$ and products thereof:
\begin{equation}
\label{connected-def}  F_{2N,c}^\ordo(\theta_1,\dots,\theta_N)\equiv
\text{(finite part of) }  F_{2N}^\ordo(\theta_1+\eps_1,\dots,\theta_N+\eps_N|\theta_N,\dots,\theta_1)
\end{equation}
The general structure of the singularities in \eqref{altalanosdiag}
was worked out in \cite{fftcsa2}. For future use we recall the
relation between the symmetric and connected evaluation schemes. Firs
we introduce the necessary notations.

Let us take $n$ vertices labeled by the numbers $1,2,\dots,n$ and
let $G$ be the set of the directed graphs $G_{i}$ with the following
properties: 
\begin{itemize}
\item $G_{i}$ is tree-like. 
\item For each vertex there is at most one outgoing edge. 
\end{itemize}
For an edge going from $i$ to $j$ we use the notation $E_{ij}$.

\begin{thm}
\label{par:Theorem-1}
The function $F^\ordo_{2N,s}(\theta_1,\dots,\theta_N)$ can be evaluated as a sum over all graphs
in $G$, where the contribution of a graph $G_{i}$ is given by the
following two rules: 
\begin{itemize}
\item Let $A_{i}=\{ a_{1},a_{2},\dots,a_{m}\}$ be the set of vertices from
which there are no outgoing edges in $G_{i}$. The form factor associated
to $G_{i}$ is \begin{equation}
F^\ordo_{2m,c}(\theta_{a_{1}},\theta_{a_{2}},\dots,\theta_{a_{m}})\label{egygrafformfaktora}\end{equation}
 
\item For each edge $E_{jk}$ the form factor above has to be multiplied
by \[\varphi(\theta_{j}-\theta_{k})\]
 \end{itemize}  
\end{thm}

This theorem was proven in \cite{fftcsa2} using the kinematical singularity
axiom \eqref{eq:kinematicalaxiom}.

\medskip

Now we are in a position to state one of the central results of
\cite{fftcsa2}, which is the following relation between
diagonal form factors in infinite and finite volume:
\begin{equation}
  \label{fftcsa2-result1}
\bra{\theta_1,\dots,\theta_N}\ordo\ket{\theta_1,\dots,\theta_N}_L=
\frac{1}{\rho_N(\theta_1,\dots,\theta_N)}
\sum_{\{\theta_+\}\cup \{\theta_-\}}
F^\ordo_{2n,s}\big(\{\theta_-\}\big)\rho_{N-n}\big(\{\theta_+\}\big)
\end{equation}
The sum runs over all possible bipartite partitions. 
Similar expansions are known for mean values in the Algebraic Bethe Ansatz literature
\cite{korepin-izergin,korepin-LL1,iz-kor-resh,korepinBook}. In that context
the symmetric evaluation is called the ``irreducible part'' of the
form factor. More details about the relation of our expansions to
those in ABA are presented in section \ref{LLsec}.

A very important result of \cite{fftcsa2} is that
\eqref{fftcsa2-result1} can be expressed alternatively in terms of the
connected evaluation of the diagonal form factors. To restate this result
we introduce a restricted Gaudin determinant as follows. For a
given bipartite partition 
\begin{equation*}
  \{\theta_1,\dots,\theta_N\} = \{\theta_+\}\cup \{\theta_-\}
\end{equation*}
\begin{equation*}
  \big|\{\theta_+\}\big|=N-n \quad\text{and}\quad \big|\{\theta_-\}\big|=n
\end{equation*}
we define the restricted determinant
\begin{equation}
\label{restricted-density}
  \bar\rho_{N-n}(\{\theta_+\}|\{\theta_-\})=\det \mathcal{J}_+
\end{equation}
where $\mathcal{J}_+$ is the sub-matrix of  $\mathcal{J}$ corresponding
to the particles in the set $\{\theta_+\}$.
Note that $\bar\rho_{N-n}(\{\theta_+\}|\{\theta_-\})$
still contains information about the complementary set of rapidities
$\{\theta_-\}$, and it should not be confused with the density
$\rho_{N-n}(\{\theta_+\})$ which depends only on the rapidities
$\{\theta_+\}$. 

With these notations, the alternative expression for the expectation
value reads
\begin{equation}
    \label{fftcsa2-result2}
\bra{\theta_1,\dots,\theta_N}\ordo\ket{\theta_1,\dots,\theta_N}_L=
\frac{1}{\rho_N(\theta_1,\dots,\theta_N)}
\sum_{\{\theta_+\}\cup \{\theta_-\}}
F^\ordo_{2n,c}\big(\{\theta_-\}\big)
\bar\rho_{N-n}\big(\{\theta_+\}|\{\theta_-\}\big)
\end{equation}
The equivalence of \eqref{fftcsa2-result1} and
\eqref{fftcsa2-result2} is proven in Theorem 2 of \cite{fftcsa2}. We
wish to note that the two relations are expected to be exact to all
orders in $1/L$, but there are residual finite size effects of
order $\ordo(e^{-\mu L})$. 
The general structure of exponential
corrections to form factors is not yet known, however it is expected
that the growth of the coefficients of the $\ordo(e^{-\mu' L})$ terms
with any $\mu'\ge \mu$ is only polynomial in
$N$. Therefore it is safe to neglect them in taking the thermodynamic
limit of  \eqref{fftcsa2-result1} and \eqref{fftcsa2-result2}.

\subsection{Evaluation of the thermodynamic limit}

\label{TMF}

We consider the diagonal matrix element
\begin{equation}
\label{omega1}
\vev{\ordo}_{N,L}=
\bra{\theta_1,\dots,\theta_N}\ordo \ket{\theta_1,\dots,\theta_N}_L
\end{equation}
with a large number of particles in a large volume $L$. The idea
is to evaluate expressions \eqref{fftcsa2-result1} and
\eqref{fftcsa2-result2} in the $L\to\infty$ limit assuming that the
distribution of roots is given by the smooth functions $\rho^{(o)}(\theta)$
and $\rho^{(h)}(\theta)$. They describe the density of occupied roots
and holes, respectively. They satisfy the constraint 
\begin{equation}
\label{Lieb}
\rho^{(o)}(\theta)+\rho^{(h)}(\theta)=\frac{p'(\theta)}{2\pi}
+\int_{-\infty}^\infty\frac{d\theta'}{2\pi} 
\varphi(\theta-\theta')\rho^{(o)}(\theta'),
\end{equation}
which follows from the thermodynamic limit of the Bethe equations. The
kernel in \eqref{Lieb} is given by $\varphi=-i \frac{d}{d\theta} \log
S(\theta)$ and $p(\theta)$ is the one-particle momentum. The total particle
density is given by the integral
\begin{equation*}
  \rho=\frac{N}{L}=\int {d\theta} \rho^{(o)}(\theta)
\end{equation*}
Let us introduce the quantities
\begin{equation*}
\begin{split}
C_{n,L}&= \sum_{\big|\{\theta_-\}\big|=n}
 F^\ordo_{2n,c}\big(\{\theta_-\}\big)
\frac{\bar\rho_{N-n}\big(\{\theta_+\}|\{\theta_-\}\big)}{\rho_N(\theta_1,\dots,\theta_N)}\\
D_{n,L}&=\sum_{\big|\{\theta_-\}\big|=n}
 F^\ordo_{2n,s}\big(\{\theta_-\}\big)
\frac{\rho_{N-n}\big(\{\theta_+\}\big)}{\rho_N(\theta_1,\dots,\theta_N)}
\end{split}
\end{equation*}
such that
\begin{equation*}
\vev{\ordo}_{N,L}=\sum_{n=0}^N  C_{n,L}=\sum_{n=0}^N  D_{n,L}
\end{equation*}
It will be shown that both $C_{n,L}$ and $D_{n,L}$ possess a
well-defined $L\to\infty$ limit for arbitrary $n$.
This way the expectation value is expressed as
\begin{equation*}
  \vev{\ordo}=\sum_{n=0}^\infty  C_{n}=\sum_{n=0}^\infty  D_{n}
\end{equation*}
where
\begin{equation*}
  C_{n}=\lim_{L\to\infty} C_{n,L}\qquad\qquad
 D_{n}=\lim_{L\to\infty} D_{n,L}
\end{equation*}
 We assume that performing the  summation
over $n$ and taking the thermodynamic limit can be
exchanged. Note that the total number of  $C_{n,L}$ and $D_{n,L}$ depends on
$L$, because $n\le [\rho L]$. However this is not a serious problem,
because for any $n$ there will be a large enough volume with $N>n$, so that
$C_{n,L}$ will appear in the sum; moreover it will have a finite limit as $L$ is
sent further to infinity. 

\bigskip

The $n=0$ term reproduces the infinite volume vacuum expectation
value:
\begin{equation*}
  C_{0,L}=D_{0,L}=\bra{0}{\ordo}\ket{0}
\end{equation*}
In the present approach this vacuum expectation value is interpreted
as the maximally disconnected part of the diagonal form factor.

In the case of  $n=1$ 
the quantities $C_{1,L}$ and $D_{1,L}$ are expressed as a single
sum over $\theta_j\in\{\theta_1,\dots,\theta_N\}$:
\begin{equation*}
\begin{split}
C_{1,L}&= \sum_{\theta_j}
 F^\ordo_{2k,c}\big(\theta_j\big)
\frac{\bar\rho_{N-1}\big(\theta_1,\dots,\hat \theta_j,\dots,\theta_N|\theta_j\big)}{\rho_N(\theta_1,\dots,\theta_N)}\\
D_{1,L}&=\sum_{\theta_j}
 F^\ordo_{2k,s}\big(\theta_j\big)
\frac{\rho_{N-1}\big(\theta_1,\dots,\hat \theta_j,\dots,\theta_N\big)}{\rho_N(\theta_1,\dots,\theta_N)}
\end{split}
\end{equation*}
Let us assume that the ratios of the determinants have a well-defined
$L\to\infty$ limit while keeping $\theta_j$ fixed. Then the summation
over $\theta_j$ can be expressed as an integral:
\begin{equation*}
  \sum_{\theta_j} \quad \to \quad \int d\theta\
  \rho^{(o)}(\theta) L
\end{equation*}
The ratios of the determinants scale as $1/L$, therefore $C_{1,L}$ and
$D_{1,L}$  behave as $\ordo(L^0)$. This is expected in order to
have a well-defined $L\to\infty$ limit. 

The limiting values of the ratios of the determinants can be
determined using the techniques of \cite{korepin-LL1,korepinBook}. In
Appendix \ref{ratioss} it is shown that
\begin{equation*}
  \begin{split}
 \lim  \frac{\bar\rho_{N-1}\big(\theta_1,\dots,\hat \theta_j,\dots,\theta_N|\theta_j\big)}
{\rho_N(\theta_1,\dots,\theta_N)}&=\frac{1}{2\pi L(\rho^{(o)}(\theta_j)+\rho^{(h)}(\theta_j))}
\\
 \lim  \frac{\rho_{N-1}\big(\theta_1,\dots,\hat \theta_j,\dots,\theta_N\big)}
{\rho_N(\theta_1,\dots,\theta_N)}&=\frac{1}{2\pi L(\rho^{(o)}(\theta_j)+\rho^{(h)}(\theta_j))} \omega(\theta_j)
  \end{split}
\end{equation*}
Here
\begin{equation*}
  \omega(\theta)=\exp\left(-\int \frac{d\theta'}{2\pi} f(\theta') \varphi(\theta-\theta')\right)
\end{equation*}
and the weight function $f(\theta)$ is defined as
\begin{equation*}
  f(\theta)
=\frac{\rho^{(o)}(\theta)}{\rho^{(o)}(\theta)+\rho^{(h)}(\theta)}
\end{equation*}
Putting everything together one obtains
\begin{equation}
\begin{split}
  \label{2p-eredm}
  C_{1}&=\lim_{L\to\infty} C_{1,L}=\int \frac{d\theta}{2\pi} f(\theta)F_{2,c}^\ordo(\theta)\\
 D_{1}&=\lim_{L\to\infty} D_{1,L}=\int \frac{d\theta}{2\pi} f(\theta)\omega(\theta)
 F_{2,s}^\ordo(\theta)
\end{split}
\end{equation}
One can repeat this line of reasoning  for the
case when a fixed ($n>1$) number of rapidities are chosen to be in the
subset $\{\theta_-\}$, which are to be substituted into a diagonal
$2n$ particle form factors. It is easy to see that in the
thermodynamic limit there will be no complications, in particular the
application of the Pauli principle for the summation over
$\{\theta_-\}$ only causes corrections of $\ordo(1/L)$. Therefore one
can immediately write down the general result
\begin{equation}
  \label{Ck}
C_n=\frac{1}{n!}\int \frac{d\theta_1}{2\pi}\dots
\frac{d\theta_n}{2\pi}\left( \prod_{j=1}^n f(\theta_j)\right)
F_{2n,c}^\ordo(\theta_1,\dots,\theta_n)
\end{equation}
and
\begin{equation}
  \label{Dk}
D_n=\frac{1}{n!}\int \frac{d\theta_1}{2\pi}\dots
\frac{d\theta_n}{2\pi} 
\left( \prod_{j=1}^n f(\theta_j)\omega(\theta_j)\right)
F_{2n,s}^\ordo(\theta_1,\dots,\theta_n)
\end{equation}
The expectation value of $\ordo$ is then given by
\begin{equation}
  \begin{split}
 \vev{\ordo}&=\sum_n \frac{1}{n!}
\int \frac{d\theta_1}{2\pi}\dots  \frac{d\theta_n}{2\pi} \left( \prod_{j=1}^n f(\theta_j)\right)
F^\ordo_{2n,c}(\theta_1,\dots,\theta_n)\\
&=\sum_k \frac{1}{n!}
\int \frac{d\theta_1}{2\pi}\dots  \frac{d\theta_n}{2\pi}
\left( \prod_{j=1}^n f(\theta_j)\omega(\theta_j)\right)
F^\ordo_{2n,s}(\theta_1,\dots,\theta_n)
\label{LMsajat}
  \end{split}
\end{equation}
The two different expressions represent a rearrangement of
terms, similar to the equivalence of \eqref{fftcsa2-result1} and
\eqref{fftcsa2-result2}. We wish to note that \eqref{LMsajat} is the
main result of this work. 

Both series in \eqref{LMsajat} are expected to be convergent. We are
not able to present a rigorous proof, because the behaviour of the
form factors is in general not known. However, it is generally assumed
that their exists a $K\in\valos^+$ such that
\begin{equation}
\label{bounded}
\big|  F^\ordo_{n+m}(\theta_1',\dots,\theta_n'|\theta_1,\dots,\theta_m)\big|< K^{n+m}\qquad\text{for}
\quad \theta_j,\theta'_j\in \valos
\end{equation}
This implies that both functions $F_{2n,s}^\ordo$ and $F_{2n,c}^\ordo$
are bounded in the sense of \eqref{bounded} and the series
\eqref{LMsajat} are convergent.

\subsection{Boundary operators in Boundary Field Theories}

\label{sec:boundary}

It is possible to extend the previous calculations to relativistic
Boundary QFT's \cite{Corrigan:1996cg,Ghoshal:1993tm}. 
Consider an operator $\ordo(t)$ localized at the boundary of a
half-infinite system. The boundary is supposed to be integrable; the
scattering off the boundary is described by the elastic reflection factor $R(\theta)$.
Asymptotic states can be defined similarly as in
the bulk case; the only difference is that for incoming (outgoing) states all the
rapidities are positive (negative). Form factors of $\ordo$ are then
defined as\footnote{Form factors of boundary operators are denoted by
  $G_N^\ordo$ to distinguish them from the form factors of bulk operators.}
\begin{equation*}
  G_N^\ordo(\theta_1,\dots,\theta_N)=\bra{0}\ordo\ket{\theta_1,\dots,\theta_N}
\end{equation*}
The analytic properties of the functions $G_N^\ordo$ (also called
boundary form factor equations) were derived in \cite{Bajnok:2006ze}. For
the sake of brevity we do not cite them here.

We will be interested in the mean values of the boundary operator $\ordo$.
We start with a finite system of size $L$
with two integrable boundaries $a,b$ with elastic reflection factors
$R_{a,b}(\theta)$; the operator $\ordo$ lives on the boundary $a$.  The Bethe equations read
\begin{equation*}
  e^{i Q^B_j}\equiv
e^{2ip_jL}R_a(\theta_j)R_b(\theta_j)\prod_{k\ne j}
S(\theta_k-\theta_j)S(\theta_k+\theta_j)=1,
\qquad \theta_j>0
\end{equation*}
The rapidities are restricted to take only positive values. The
$N$-particle density of states is obtained as
\begin{equation*}
 \rho^B_N(\{\theta\})=\det \mathcal{J}^B,\qquad
 \mathcal{J}^B_{ij}=\frac{\partial Q^B_i}{\partial \theta_j}
\end{equation*}
Analogously to \eqref{restricted-density} we define a restricted
density for a given partition
$\{\theta\}=\{\theta_+\}\cup\{\theta_-\}$ as the sub-determinant
belonging to the subset $\{\theta_+\}$:
\begin{equation*}
   \bar\rho^B_N(\{\theta_+\},\{\theta_-\})=\det \mathcal{J}^B_+
\end{equation*}

The relation between finite volume and infinite
volume form factors was worked out in \cite{Kormos:2007qx}. 
In the generic case with no coinciding rapidities they take the same form
as \eqref{FVFF} with the obvious replacements $F_N^\ordo\to G_N^\ordo$
and $\rho_N\to\rho^B_N$. For diagonal form factors one obtains
\begin{equation}
    \label{fftcsa2-boundary}
\bra{\theta_1,\dots,\theta_N}\ordo\ket{\theta_1,\dots,\theta_N}_L=
\frac{1}{\rho^B_N(\theta_1,\dots,\theta_N)}
\sum_{\{\theta_+\}\cup \{\theta_-\}}
G^\ordo_{2n,c}\big(\{\theta_-\}\big)
\bar\rho^B_{N-n}\big(\{\theta_+\}|\{\theta_-\}\big),
\end{equation}
where $G^\ordo_{2n,c}$ are the connected parts of the diagonal form
factors defined analogously to \eqref{connected-def}. Clearly,
\eqref{fftcsa2-boundary} is a simple generalization of the relation in
the bulk 
\eqref{fftcsa2-result2}. It is also possible to define the symmetric
evaluation of the boundary form factors as in \eqref{symmetric-def}, but
 the naive generalization of
\eqref{fftcsa2-result1} does not hold in the boundary case, as explained
in detail in \cite{Kormos:2007qx}.

We are interested in the thermodynamic limit of
\eqref{fftcsa2-boundary} along the lines of the previous
subsection. Using the techniques of Appendix \ref{ratioss}
it can shown that the behaviour of the ratios of determinants is the
same as in the bulk. For example 
\begin{equation*}
\lim \frac{\bar\rho^B_N(\theta_1,\dots,\hat \theta_j,\dots,\theta_N|\theta_j)}
{\rho^B_N(\theta_1,\dots,\theta_N)}=\frac{1}{2\pi L (\rho^{(o)}(\theta)+\rho^{(h)}(\theta))},
\end{equation*}
where $\rho^{(o)}(\theta)$ and $\rho^{(h)}(\theta)$ describe the
densities of occupied roots and holes, respectively. Therefore, the
thermodynamic limit of \eqref{fftcsa2-boundary} results in the series
\begin{equation}
  \begin{split}
 \vev{\ordo}&=\sum_n \frac{1}{n!}
\int_0^\infty \frac{d\theta_1}{2\pi}\dots
\int_0^\infty\frac{d\theta_n}{2\pi} 
\left( \prod_{j=1}^n f(\theta_j)\right)
G^\ordo_{2n,c}(\theta_1,\dots,\theta_n),
\label{LMboundary}
  \end{split}
\end{equation}
where
\begin{equation*}
  f(\theta)=\frac{\rho^{(o)}(\theta)}{\rho^{(o)}(\theta)+\rho^{(h)}(\theta)}
\end{equation*}

In the case of finite temperature expectation values one can show that
the distribution of roots will be given by the same functions as in the periodic
case, as expected by thermodynamic arguments. Therefore 
\begin{equation}
  \begin{split}
 \vev{\ordo}_{T,\mu}&=\sum_n \frac{1}{n!}
\int_0^\infty \frac{d\theta_1}{2\pi}\dots
\int_0^\infty\frac{d\theta_n}{2\pi} 
\left( \prod_{j=1}^n \frac{1}{1+e^{\eps(\theta_j)}}\right)
G^\ordo_{2n,c}(\theta_1,\dots,\theta_n),
\label{LMboundary2}
  \end{split}
\end{equation}
where $\eps(\theta)$ is the solution of the TBA with periodic boundary
conditions \eqref{TBA}. The series \eqref{LMboundary2} was proposed 
in \cite{Takacs:2008ec} and it was proven there up to third
non-trivial order using the techniques
of \cite{fftcsa2}.
The all-orders proof
together with the general formula \eqref{LMboundary} is a new result
of this work.

\section{Application to the 1D Bose gas}

\label{LLsec}

In this section we consider the 1D interacting Bose gas also known as
the Lieb-Liniger (LL)
model \cite{Lieb-Liniger,korepinBook}. 
The second quantized form of the Hamiltonian in volume $L$ with periodic boundary
conditions is given by
\begin{equation}
\label{H-LL}
H_{\text{LL}}=\int_{0}^{L}\,\mathrm d x\left(\p_x\Psi^\dagger\p_x\Psi+ c\Psi^\dagger\Psi^\dagger\Psi\Psi\right)\;.
\end{equation}
Here $\Psi(x,t)$ and $\Psi^\dagger(x,t)$ are  canonical
non-relativistic Bose
fields satisfying
\begin{equation}
  [\Psi(x,t),\Psi^\dagger(y,t)]=\delta(x-y)\;,
\end{equation}
and  $c$ is the coupling constant. We used the conventions  $m=1/2$ and $\hbar=1$.
The Fock vacuum is defined as 
\begin{equation*}
  \Psi\vac=0,\qquad\cav\Psid=0\;
\end{equation*}
The eigenstates of the Hamiltonian \eqref{H-LL} can be constructed
using the Bethe Ansatz. The scattering of the particles is described by
the two-particle S-matrix
\begin{equation*}
  S_{LL}=\frac{\lambda-ic}{\lambda+ic},
\end{equation*}
where $\lambda$ is the non-relativistic rapidity variable and
multi-particle energies and momenta are given by
\begin{equation*}
  E_N=\sum_j \lambda_j^2\qquad\qquad P_N=\sum_j \lambda_j
\end{equation*}

The Algebraic Bethe Ansatz (ABA) provides a framework to calculate form factors and correlation
functions. However, before turning to the ABA solution we present the
results of the papers \cite{sinhG-LL1,sinhG-LL2}, where a
non-relativistic limit of the sinh-Gordon theory was performed to
obtain physical quantities in the LL model. They
considered the temperature dependent expectation values
\begin{equation*}
  \vev{\ordo_k}\equiv\vev{{\Psi^\dagger}^k\Psi^k}=g_k(T,c)\ \rho^k
\end{equation*}
where $\rho=N/L$ is the particle
density. The dimensionless quantities $g_k$ are important
for the phenomenology of the Bose gas, for example $g_3$ describes
the recombination rate of the gas \cite{g3,low-D-trapped}. 

The main result of  \cite{sinhG-LL1,sinhG-LL2} is the integral series
\begin{equation}
\label{LM-LL}
   \vev{\ordo_k}=
\sum_{N=k}^\infty \frac{1}{N!}
\int \frac{d\lambda_1}{2\pi}\dots  \frac{d\lambda_N}{2\pi}
 \left( \prod_{j=1}^N \frac{1}{1+e^{\eps(\lambda_j)}}\right)
F^k_{2N,c}(\lambda_1,\dots,\lambda_N)
\end{equation}
where $\eps(\lambda)$ is the solution of the non-relativistic TBA equation
\begin{equation}
\label{LL-TBA}
   T \eps(\theta)=\lambda^2 -\mu-T
  \int_{-\infty}^\infty\frac{d\theta'}{2\pi} 
\varphi(\theta-\theta') \log(1+e^{-\eps(\theta')}),
\end{equation}
where 
\begin{equation*}
  \varphi(\lambda)=-i\frac{\partial }{\partial \lambda} \log S_{LL}=
\frac{2c}{c^2+\lambda^2}
\end{equation*}
The functions $F^k_{2N,c}$  appearing in \eqref{LM-LL} were derived by the non-relativistic limit of
certain form factors in the sinh-Gordon model. The first few examples
were given explicitly in \cite{sinhG-LL2}. They vanish for $N<k$ and the asymptotic behaviour
at $c\to\infty$ is given by
\begin{equation*}
  F^k_{2N,c}\sim c^{-(k(k-2)+N)}
\end{equation*}

In the following we re-establish the series  \eqref{LM-LL} using the
methods of Algebraic Bethe Ansatz
\cite{korepin-izergin,korepin-LL1,iz-kor-resh,korepinBook}. Also, we
clarify the connection between the functions 
$F^k_{2N,c}$ above and the form factors calculated from ABA.

\bigskip

In the ABA approach the central object is the monodromy matrix, which
can be written as
\begin{equation}
  T(\lambda)=
  \begin{pmatrix}
    A(\lambda) & B(\lambda)\\
C(\lambda) & D(\lambda)
  \end{pmatrix}
\end{equation}
The commutation relations satisfied by the entries can be expressed in
a compact form as
\begin{equation}
\label{RTT}
  R(\lambda,\mu)T(\lambda)\otimes T(\mu)=T(\mu)\otimes T(\lambda) R(\lambda,\mu)
\end{equation}
Here $ R(\lambda,\mu)$ the R-matrix is of XXX-type:
\begin{equation*}
  R(\lambda,\mu)=
  \begin{pmatrix}
    f(\lambda,\mu) & 0  &0  &0 \\
0 & g(\lambda,\mu) & 1 & 0\\
0 & 1 & g(\lambda,\mu) & 0\\
0 & 0 & 0 &   f(\lambda,\mu)
  \end{pmatrix}
\end{equation*}
where
\begin{equation*}
  f(\lambda,\mu)=\frac{\lambda-\mu+ic}{\lambda-\mu}\qquad\qquad 
 g(\lambda,\mu)=\frac{ic}{\lambda-\mu}
\end{equation*}
The vacuum eigenvalues of the operators $A(\lambda)$, $C(\lambda)$ and $D(\lambda)$
are
\begin{equation*}
C(\lambda)\ket{0}=0\qquad A(\lambda)\ket{0}=a(\lambda) \ket{0}\qquad
D(\lambda)\ket{0}=d(\lambda)\ket{0}  
\end{equation*}
For definiteness we specify
\begin{equation*}
  a(\lambda)=e^{-i\lambda L/2}\qquad   d(\lambda)=e^{i\lambda L/2}
\end{equation*}
The function $l(\lambda)=a(\lambda)/d(\lambda)=e^{-i\lambda L}$ will
be used extensively.

The normalized operators $ \ABAB(\lambda)$ and $
\ABAC(\lambda)$ are given by
\begin{equation*}
  \ABAB(\lambda)=B(\lambda)/d(\lambda)\qquad
 \ABAC(\lambda)=C(\lambda)/d(\lambda)
\end{equation*}
The Bethe states are then defined as
\begin{equation}
\label{ABA-states}
  \ket{\lambda_1,\dots,\lambda_N}_L=\prod_{j=1}^N \ABAB(\lambda_j)\ket{0}
\end{equation}
These states are eigenvectors of the Hamiltonian \eqref{H-LL} if the rapidities
satisfy the Bethe equations:
\begin{equation}
\label{BE-ABA}
  l(\lambda_j)\prod_{k\ne j} S_{LL}(\lambda_k-\lambda_j)=1 
\end{equation}
For later use we also introduce the logarithm of the l.h.s. as
\begin{equation*}
  Q_j(\lambda)=i \log l(\lambda)+i\sum_{k\ne j}\log
  S_{LL}(\lambda-\lambda_k)=
\lambda L +2\sum_{k\ne j} \text{atan}\frac{c}{\lambda_j-\lambda_k}
\end{equation*}

We will be interested in the matrix elements of the operators
\begin{equation*}
  {\ordo}^k=(\Psi^\dagger)^k \Psi^k
\end{equation*}
First of all let us define
the un-normalized ``off-shell'' matrix element as
\begin{equation}
\label{Mdef}
M^k_{2N}(\{\lambda^C\}_N,\{\lambda^B\}_N,\{l^C\}_N,\{l^B\}_N)=
\bra{0}\prod_{j=1}^N \ABAC(\lambda_j^C)(\Psi^\dagger)^k \Psi^k 
\prod_{k=1}^N \ABAB(\lambda_k^B)\ket{0}  
\end{equation}
Here it is understood that the r.h.s. is calculated using
the commutation relations \eqref{RTT} only; the Bethe equations
\eqref{BE-ABA} are not assumed. This way the function $M^k_N$ depends
on $4N$ variables given by $\lambda_j^C$, $\lambda^B_j$ and
$l^C_j=l(\lambda_j^C)$, $l^B_j=l(\lambda_j^B)$.  Note that
$M^k_{2N}=0$ if $k>N$.

The ``on-shell'' matrix element or ``form factor'' is defined as a
special case of \eqref{Mdef} when the two sets of rapidities are
solutions to the Bethe equations. In this case the variables $l_j$
can be expressed in terms of the functions
$f(\lambda_i,\lambda_j)$ and one obtains the function 
\begin{equation*}
  \mathbb{F}^k_{2N}(\{\lambda^C\}_N|\{\lambda^B\}_N)=
M^k_{2N}(\{\lambda^C\}_N,\{\lambda^B\}_N,\{l^C\}_N,\{l^B\}_N)
\end{equation*}
which only depends on $2N$ variables.

It can be shown that the form factor has first order poles in the case
of colliding rapidities. The residue at $\lambda^C_N\to \lambda_N^B$  is
given by \cite{iz-kor-resh,PhysRevD.23.3081}
\begin{equation}
\label{armin}
  \begin{split}
    &  \mathbb{F}^k_{2N}(\{\lambda^C\}_N|\{\lambda^B\}_N)|_{\lambda^C_N\to \lambda_N^B}
\to\\& \frac{ic}{\lambda_N^C-\lambda_N^B}
\left(\prod_{j=1}^{N-1} f_{jN}^C f_{Nj}^B -
\prod_{j=1}^{N-1} f_{jN}^B f_{Nj}^C
\right)\times
\mathbb{F}^k_{2N-2}(\{\lambda^C\}_{N-1}|\{\lambda^B\}_{N-1})
  \end{split}
\end{equation}
Here $f_{jN}^{B,C}=f(\lambda_j^{B,C}-\lambda_N^{B,C})$.
Other cases $\lambda^C_i\to \lambda_j^B$ follow from the symmetry properties of the form
factor. Note that equation \eqref{armin} does not depend on $k$,
ie. the singularity properties are the same for all
$\ordo_k$. Moreover, it can be shown that an analogous equation holds for
the operators
\begin{equation*}
  \ordo_{k,l}=(\Psi^\dagger)^k \Psi^l,\qquad k,l\in\mathbb{N}
\end{equation*}
In Appendix \ref{sec:ABAproofs} we present a general proof of \eqref{armin}.

Note that \eqref{armin} has essentially the same structure as the
kinematic singularity axiom \eqref{eq:kinematicalaxiom} in relativistic
QFT. There are two differences which are related to different
normalizations of the Bethe vectors; this was pointed out in \cite{nonrelFF}.
First of all, the norm of the states \eqref{ABA-states} is equal to
\begin{equation}
\label{normalis}
  \skalarszorzat{\lambda_1,\dots,\lambda_N}{\lambda_1,\dots,\lambda_N}=
c^N \Big(\prod_{j\ne k} f_{jk}\Big) \rho_N(\lambda_1,\dots,\lambda_N)
\end{equation}
Here $\rho_N(\lambda_1,\dots,\lambda_N)$ is the Gaudin determinant:
\begin{equation*}
  \rho_N(\lambda_1,\dots,\lambda_N)=\det \mathcal{J}, \qquad\qquad
  \mathcal{J}_{jk}=\frac{\partial Q_j}{\partial \lambda_k}
\end{equation*}
On the other hand, in relativistic situation the norm of the finite
volume states is given simply by the density of states $\rho_N$, as it
is implicitly assumed in \eqref{FVFF}. 

The other issue is related to
the exchange of rapidities in the Bethe state. The $B(\lambda)$
operators commute with each other, therefore the states
\eqref{ABA-states} are totally symmetric with respect to the exchange
of rapidities. This property also applies to the finite volume matrix
elements in relativistic QFT, as it was emphasized in
\cite{fftcsa1}. On the other hand, in relativistic QFT the infinite volume 
form factors satisfy the exchange property \eqref{eq:exchangeaxiom}, which
follows from the Faddeev-Zamolodchikov algebra. In the present case
the S-matrix is a pure phase, therefore one can introduce the phase
structure by multiplying with factors of $\sqrt{S(\lambda)}$. With a
slight abuse of notation the
differences between the normalization conventions can be 
summarized as \cite{nonrelFF}
\begin{equation}
  \ket{\lambda_1,\dots,\lambda_N}_{ABA}\qquad \sim \qquad c^{N/2} \Big(\prod_{j< k}^N f_{jk}\Big)
  \ket{\lambda_1,\dots,\lambda_N}_{QFT}
\end{equation}
In order to make contact with the form factors appearing in
the series \eqref{LM-LL} we define
\begin{equation}
\label{ujnorma}
 {F}^k_{2N}(\{\lambda^C\}_N|\{\lambda^B\}_N)=
 \Big(c^N\prod_{j< k}^N f_{jk}^Bf_{jk}^C\Big)^{-1}\
   \mathbb{F}^k_{2N}(\{\lambda^C\}_N|\{\lambda^B\}_N)
\end{equation}
The functions ${F}^k_{2N}$ satisfy the exchange property \eqref{eq:exchangeaxiom}
and the non-relativistic version of the kinematical singularity axiom \eqref{eq:kinematicalaxiom}.

Now we are in a position to define the diagonal limits of the
non-relativistic form factors. In accordance with
\eqref{symmetric-def} and \eqref{connected-def} we introduce
\begin{equation}
\label{symmdefLL}
F^k_{2N,s}(\lambda_1,\dots,\lambda_N)\equiv
\lim_{\eps\to 0}  F_{2N}^k(\lambda_1+\eps,\dots,\lambda_N+\eps|\lambda_N,\dots,\lambda_1)
\end{equation}
\begin{equation}
\label{conndefLL}
  F^k_{2N,c}(\lambda_1,\dots,\lambda_N)\equiv
\text{(finite part of) }  F_{2N}^k(\lambda_1+\eps_1,\dots,\lambda_N+\eps_N|\lambda_N,\dots,\lambda_1)
\end{equation}
The properties of these functions are the same as in the relativistic
case. Both are completely symmetric in their
variables and the relation between $F^k_{2N,s}$ and $F^k_{2N,c}$ is
given by Theorem \ref{par:Theorem-1}. We remark that the normalization
in \eqref{ujnorma} commutes with the diagonal limit, because the extra
factors do not introduce any new poles.

Now we turn our attention to the mean values of the operators
$\ordo_k$. One can define the diagonal limit of \eqref{Mdef} as
\begin{equation}
\label{Mdef2}
 M^k_N(\{\lambda\}_N,\{l\}_N,\{z\}_N)=
\lim_{\lambda^C_j\to \lambda_j^B}  M^k_N(\{\lambda^C\}_N,\{\lambda^B\}_N,\{l^C\}_N,\{l^B\}_N)
\end{equation}
Formally it will depend on $3N$ independent variables
$\{\lambda\}_N$, $\{l\}_N$ and $\{z\}_N$, where 
\begin{equation*}
  z(\lambda)=i\frac{d}{d\lambda}\log  l(\lambda)
\end{equation*}
The expectation value is then given by the special case of
\eqref{Mdef2} when the rapidities satisfy the Bethe equations;
the remaining task is to express it in terms
of the form factors defined in \eqref{symmdefLL} and
\eqref{conndefLL}. The result for the normalized expectation value is given by the following formula:
\begin{equation}
  \label{fftcsa2-result-LL}
\begin{split}
\vev{\mathcal{\ordo}_k}_N&=
\frac{1}{\rho_N(\lambda_1,\dots,\lambda_N)}
\sum_{\{\lambda_+\}\cup \{\lambda_-\}}
F^k_{2n,s}\big(\{\lambda_-\}\big)\rho_{N-n}\big(\{\lambda_+\}\big)\\
&=
\frac{1}{\rho_N(\lambda_1,\dots,\lambda_N)}
\sum_{\{\lambda_+\}\cup \{\lambda_-\}}
F^k_{2n,c}\big(\{\lambda_-\}\big)
\bar\rho_{N-n}\big(\{\lambda_+\}|\{\lambda_-\}\big)
\end{split}
\end{equation}
Clearly, the above formulas are the non-relativistic versions of
\eqref{fftcsa2-result1} and \eqref{fftcsa2-result2}. The equivalence
between the first and the second line can be proven with the
non-relativistic version of Theorem 2 in \cite{fftcsa2}.
Although \eqref{fftcsa2-result-LL}
could be proven by performing a non-relativistic limit along the lines of
\cite{sinhG-LL1,sinhG-LL2}, we felt it worthwhile to provide a
derivation using only the techniques of ABA. This is presented in Appendix
\ref{sec:mv}.

Performing the thermodynamic limit of \eqref{fftcsa2-result-LL} along the lines of subsection
\eqref{TMF} one readily derives the integral series \eqref{LM-LL}
and also the alternative form
\begin{equation}
\label{LM-LL2}
   \vev{\ordo_k}=
\sum_N \frac{1}{N!}
\int \frac{d\lambda_1}{2\pi}\dots  \frac{d\lambda_N}{2\pi}
 \left( \prod_{j=1}^N \frac{\omega(\lambda_j)}{1+e^{\eps(\lambda_j)}}\right)
F^k_{2N,s}(\lambda_1,\dots,\lambda_N)
\end{equation}
where
\begin{equation*}
  \omega(\lambda)=\exp\left(-\int \frac{d\lambda'}{2\pi} 
\frac{1}{1+e^{\eps(\lambda')}} \varphi(\lambda-\lambda')\right)
\end{equation*}
Equation \eqref{LM-LL2} is a new result of this work. It is
reminiscent of the integral series for the two-point functions derived
in \cite{korepin-izergin,korepin-LL1,iz-kor-resh}. In fact these works
consider the
thermodynamic expectation value
\begin{equation*}
  \vev{e^{\alpha Q(x)}},\quad \alpha \in \mathbb{C},
\end{equation*}
where
\begin{equation*}
 Q(x)=\int_0^x \Psi^\dagger(x)\Psi(x)
\end{equation*}
Expectation values of (powers of) $Q(x)$ are then obtained as
\begin{equation}
 \left. \frac{\partial}{\partial \alpha}  e^{\alpha Q(x)}\right|_{\alpha=0}=Q(x),
\qquad\qquad
 \left. \frac{\partial^2}{\partial \alpha^2}  e^{\alpha Q(x)}\right|_{\alpha=0}=Q^2(x).
\end{equation}
The two-point functions are obtained further by differentiating
w.r.t. $x$. For example the current-current correlator is given by
\begin{equation}
\label{JJ}
  \vev{J(x)J(0)}=-\frac{1}{2}\frac{d^2}{dx^2}Q^2(x),\quad\qquad J(x)=\Psi^\dagger(x)\Psi(x)
\end{equation}
In \cite{korepin-izergin,korepin-LL1} a form factor expansion was
derived for the above correlator. The resulting series has similar
properties to our formula \eqref{LM-LL2}, in particular the weight functions are
exactly the same. On the other hand, the two-point function is a
more complicated object because its $x$-dependence, and the integral series in
\cite{korepin-izergin,korepin-LL1} involve a non-trivial dressing
procedure for the momenta of the multi-particle excitations.

It seems natural that our series \eqref{LM-LL2} could be obtained by
taking the (properly regularized) $x\to 0$ limit of the expectation
values of $Q^k(x)$. In Appendix \ref{sec:xto0} we show how this limit
works for the finite volume matrix elements \eqref{fftcsa2-result-LL} in the case of
$k=2$ and 
 $N=2$. In principle the $x\to 0$
limit could be performed for arbitrary $k$ and $N$, then the thermodynamic
limit would follow in a straightforward way.
However, we do not pursue this problem here, because the intention of this work was
to derive the results for the one-point functions in a direct way.

In order to evaluate the series \eqref{LM-LL}-\eqref{LM-LL2}
one has to determine the form factors $F_{2N,c}^k$ and
$F_{2N,s}^k$. The first few cases were derived in
\cite{sinhG-LL1,sinhG-LL2} using the non-relativistic limit of the
sinh-Gordon form factors. In principle this procedure can be performed
for any $k$ and $N$, moreover it is easy to  implement
it with symbolic manipulation programs. It was demonstrated in \cite{sinhG-LL1,sinhG-LL2} that in the strong
coupling regime it is sufficient to consider only the
first few terms in the integral series.
However the general forms of  $F_{2N,c}^k$ and
$F_{2N,s}^k$   are 
not know, therefore at present the exact result
\eqref{LM-LL}-\eqref{LM-LL2} can be considered a formal expansion.

In Appendix
\ref{sec:FFs} we derive the form factors for $k=1,2$ and arbitrary
$N$; this gives an indepentent confirmation of the results of
\cite{sinhG-LL1,sinhG-LL2} in the cases $k=1,2$ and $N=1,2,3$.
 On the other hand, the general results are not known for $k\ge
3$. One possibility is to solve the recursive equations \eqref{armin} or to
employ the techniques of \cite{maillet-inverse-scatt,higher-spin-XXX-FF} to obtain
determinant formulas for the off-diagonal form factors; the diagonal limit should
be taken afterwards. However, this problem is beyond the scope of the present
work. 

\section{Application to quench problems}

\label{sec:quench}

In \cite{davide} the authors considered the real-time evolution of the
expectation value of a local operator in integrable QFT after a certain type of
quench, which changes the Hamiltonian from $H_0$
to $H$, where $H$ (possibly also $H_0$) is considered to be integrable.
The main assumption is that the initial state of
the system (which is the ground state of $H_0$) can be expanded in the
multi-particle basis of the integrable Hamiltonian $H$ in the form of  a boundary
state \cite{Ghoshal:1993tm}. In the simplest case with only one
particle type in the spectrum the corresponding expression is 
\begin{equation*}
  \ket{B}=
 \exp\left(\int \frac{d\theta}{4\pi}
      K(\theta) A(-\theta)A(\theta) \right) \ket{0}\;,
\end{equation*}
where $K(\theta)$ is an arbitrary function satisfying
$K(\theta)=S(2\theta)K(-\theta)$. The time evolution of an expectation
value is then given by
\begin{equation*}
  \vev{\ordo(0,t)}=\bra{B}e^{iHt}\ordo(0,0)e^{-iHt}\ket{B}
\end{equation*}
The main idea is that in the $t\to\infty$ limit (and taking a proper time
average) only the diagonal
matrix elements contribute and the off-diagonal ones can be
neglected. The authors then arrive at the final expression
\begin{equation}
\label{fioretto}
  \vev{\ordo}=\lim_{T\to\infty} \frac{1}{T}\int_0^T dt \vev{\ordo(0,t)}=
\sum_{n=0}^\infty \frac{1}{n!}
\prod_i \left\{\int \frac{d\theta_i}{2\pi}
\frac{|G(\theta_i)|^2}{1+|G(\theta_i)|^2}
\right\}
F_{2n,c}^\ordo(\theta_1,\dots,\theta_n)\;,
\end{equation}
where the weight function is given by the solution of the TBA-like equation
\begin{equation}
\label{quench-TBA}
- \log  |G(\theta)|^2=-\log |G_0(\theta)|^2
-\int \frac{d\theta'}{2\pi}\varphi(\theta-\theta')    \log\big(1+|G(\theta')|^2\big)
\end{equation}
The source term is given by
\begin{equation*}
G_0=e^{-2m\tau_0\cosh\theta}K(\theta),
\end{equation*}
where $\tau_0$ is a UV cut-off, which can be interpreted as the
extrapolation length known in the theory of critical boundary
phenomena \cite{Diehl-surface-QFT}. From a formal
point of view a finite $\tau_0$ is needed to have a
normalizable boundary state. 

Here we re-derive equation \eqref{fioretto} from our general result
\eqref{Apollo11}. It will be shown that \eqref{Apollo11} is applicable with the weight function
\begin{equation*}
  f(\theta)=\frac{|G(\theta)|^2}{1+|G(\theta)|^2}
\end{equation*}
First we consider a large volume $L$. 
In finite volume the boundary state is given by \cite{Kormos:2010ae}
\begin{equation*}
  \ket{B}_L=\sum_{N=0}^\infty \sum_{\theta_1,\dots,\theta_N}
  \mathcal{N}(\theta_1,\dots,\theta_N)
 K(\theta_1)\dots K(\theta_N) \ket{-\theta_1,\theta_1,\dots,-\theta_N,\theta_N}_L
\end{equation*}
Here the summation runs over all parity symmetric configurations,
where the set $\{-\theta_1,\theta_1,\dots,-\theta_N,\theta_N\}$ is
assumed to satisfy the $2N$-particle Bethe equations and we require
$\theta_i>\theta_j$ for $i>j$. The additional
normalization factors $\mathcal{N}(\theta_1,\dots,\theta_N)$ are
specific to the finite volume situation. They are given by  
\begin{equation*}
   \mathcal{N}(\theta_1,\dots,\theta_N)=
\frac{\sqrt{\rho_{2N}(-\theta_1,\theta_1,\dots,-\theta_N,\theta_N)}}{\tilde\rho_N(\theta_1,\dots,\theta_N)},
\end{equation*}
where $\rho_{2N}$ is the usual $2N$-particle density of states, and
$\tilde\rho_N$ is the constrained density in the space of rapidity
pairs \cite{Kormos:2010ae}. It can be shown that in the thermodynamic
limit $\mathcal{N}$ will converge to a finite value \cite{sajat-g},
and it will drop out from the calculation of the expectation value.

The boundary state is normalizable with any finite $\tau_0$ and the
(time averaged) expectation value is given by
\begin{equation}
\vev{\ordo}=
\lim_{L\to\infty}\lim_{T\to\infty}\frac{1}{T}
\int_{0}^T dt\ \frac{_L\bra{B}e^{iH_Lt}\ordo(0,0)e^{-iH_Lt}\ket{B}_L}{_L\skalarszorzat{B}{B}_L}
\label{limits}
\end{equation}
We assume that in the large time limit it is sufficient to consider
only the diagonal matrix elements.
In this case the calculation boils down to a simple statistical
average:
\begin{equation}
 \vev{\ordo}
=\lim_{L\to\infty}\frac{\sum_i |w_i|^2 \bra{i}\ordo\ket{i}_L}{\sum_i |w_i|^2}
\label{errevarrjgombot}
\end{equation}
where for simplicity we denoted
\begin{equation*}
\begin{split}
  \ket{i}_L=\ket{-\theta_1,\theta_1,\dots,-\theta_N,\theta_N}_L\qquad\qquad
w_i= \mathcal{N}(\theta_1,\dots,\theta_N)
 K(\theta_1)\dots K(\theta_N)
\end{split}
\end{equation*}
This statistical average \eqref{errevarrjgombot} can be considered as
a grand-canonical ensemble for particle pairs $(-\theta,\theta)$. 
In fact, one can identify the weights as
\begin{equation*}
  |G_0(\theta)|^2\sim e^{-E(\theta)}
\end{equation*}
where we set $T=1$ and $E(\theta)$ is interpreted as a bare excitation
energy: 
\begin{equation}
\label{bare}
  E(\theta)=4m\tau_0\cosh\theta -\log |K(\theta)|^2
\end{equation}
For any regular $K(\theta)$ this bare energy has the typical features
of a kinetic term: it is a regular function for a 
small $\theta$, whereas for large $|\theta|$ it goes to infinity.

One can now apply standard arguments to look for those configurations
which dominate the average \eqref{errevarrjgombot}. Repeating all the
steps which lead to the Boundary TBA equations \cite{LeClair:1995uf} it is then a standard
exercise to arrive at \eqref{quench-TBA}. In
fact, \eqref{quench-TBA} is identical to the BTBA equation of
\cite{LeClair:1995uf} with the substitution $R=\tau_0$. 

With this we have shown that the distribution of roots is governed by 
\eqref{quench-TBA}, therefore \eqref{fioretto} follows from our general
formula \eqref{Apollo11}. Let us make some further remarks about
this result. 

The most remarkable property of \eqref{fioretto} was already pointed
out in \cite{davide}: although the boundary states only include
rapidity pairs, the form factors in the final result represent
processes with single particle excitations. This can be understood by
recalling how we derived our main formulas in subsection
\ref{TMF}. There it was shown that the individual contributions of
the LeClair-Mussardo formula arise from disconnected terms of a matrix
element with a large number of particles. Moreover, the thermodynamic
limit of the mean value does not depend on the details of the finite
volume state, but only on the distribution of roots. Therefore it is
completely irrelevant that the actual state only involves particle-pairs.
The distribution of roots is always parity symmetric, therefore the
final result is completely consistent with the
structure of the boundary state.

\bigskip

To conclude this section we discuss the implications of the main
result \eqref{LMsajat} for more general quench situations.
Consider the time-evolution generated by an
integrable Hamiltonian $H$ from an initial state $\Psi_0$, which does
not have the form of a boundary state.
Neglecting possible complications due to degeneracies,
the infinite-time limit of local observables will be given
by the Diagonal Ensemble
\begin{equation}
 \vev{\ordo}=
\frac{\sum_\alpha |c_\alpha|^2 \bra{\alpha}\ordo\ket{\alpha}_L}{\sum_\alpha |c_\alpha|^2},
\label{DE}
\end{equation}
where the summation runs over the eigenstates of $H$ and 
\begin{equation*}
  c_\alpha=\skalarszorzat{\alpha}{\Psi_0}
\end{equation*}
Typically the overlaps are not known, therefore it is very hard to
determine which states will have an important effect on the
average \eqref{DE}. On the other hand, in the thermodynamic limit it is natural
to assume that the dominant states will have a smooth distribution of
roots.

Consider such a state $\ket{\alpha}$ and the mean value $\bra{\alpha}\ordo\ket{\alpha}_L$.
According to \eqref{LMsajat}
it always  takes the form of a thermal average with 
the weight functions determined by the root densities.
Assuming
that there is a one-to-one correspondence between the 
distributions and the infinite set of conserved charges this means
that the mean value only depends on the macroscopic value of the
conserved quantities and not on the details of
the state $\alpha$.
 This result is a generalized form of the ``Eigenstate Thermalization
Hypothesis''  \cite{PhysRevA.43.2046,1994PhRvE..50..888S} appropriate
to integrable theories \cite{2008Natur.452..854R}. 
 Assuming further that the dominant states in
\eqref{DE} will have the same set of macroscopic conserved charges we conclude
that the average \eqref{DE} can be substituted by a single mean value
$\bra{\alpha}\ordo\ket{\alpha}_L$ and then the integral series
\eqref{LMsajat} applies. This result can be interpreted
as a Generalized Gibbs
Ensemble (GGE) as proposed in \cite{rigol-gge}, although 
 the relation
to the conserved charges is rather indirect.

In order to apply these results in other quench situations the
assumptions about the root distributions have to be justified. More
specifically it needs to
be checked whether all relevant states in  \eqref{DE} can be described
by a single distribution. This can be a quite
challenging problem. We wish to stress that our approach only applies to
the thermodynamic limit with the prescription \eqref{limits}
and it is not clear how to obtain finite size
effects or observables at large but not infinite times with the
present methods \footnote{In the problem of Fioretto and Mussardo
  expectation values at finite times can be obtained using the methods
of \cite{Kormos:2010ae}. However, so far only the
first few terms have been calculated
explicitly and the general all-orders result is not yet available.}.

\section{Conclusions}

\label{sec:conclusions}

In this paper we studied mean values of local operators in Bethe
states with a large number of particles and a smooth distribution of
roots. The main result is formula \eqref{LMsajat}; it can be
considered as a generalization of the LeClair-Mussardo formula originally
proposed
in \cite{leclair_mussardo}. We showed that it
applies both to relativistic field theory and the non-relativistic
Lieb-Liniger model. The individual terms in the series arise from
disconnected pieces of the diagonal matrix element; the $m$th term
represents $m$ particle processes over the Fock-vacuum. 

Our results are analogous to the expansion for the two-point
functions in the Bose gas derived in the papers
\cite{korepin-izergin,korepin-LL1,iz-kor-resh}. The finite volume
formulas \eqref{fftcsa2-result-LL} are a new result of this work and
they serve as a starting point for the thermodynamic
limit. The functions $F_{2m,s}^k$ in the first line of \eqref{fftcsa2-result-LL} (the
symmetric evaluations of the diagonal form factors) are analogous to
the ``irreducible parts'' of the correlators known in the ABA
literature. On the other hand, the expansion in the second line of \eqref{fftcsa2-result-LL}
employs the functions  $F_{2m,c}^k$ which are the connected parts of
the infinite volume diagonal form factors. Relativistic Field Theory
motivated the use of the connected parts and they seem to be new
in the context of Algebraic Bethe Ansatz.

It would be interesting to extend the present approach to non-diagonal
scattering theories, which correspond to nested Bethe Ansatz
systems. In integrable Field Theory the infinite volume form factors are explicitly
known in a number of models \cite{smirnov_ff,Babujian:2006km}, but the
relation to the finite volume form factors has not yet been
determined. The first step would be to write down the corresponding
generalization of \eqref{fftcsa2-result1} for the finite volume mean
values. If such an expression is found, then the thermodynamic limit
could yield a LeClair-Mussardo formula for non-diagonal scattering
theories. On the other hand, it is not clear how much can be achieved
in non-relativistic (nested) Bethe Ansatz systems. At present the only
available results concern the spectrum (including the thermodynamics
\cite{Takahashi-book}) or the norms of Bethe wave functions
\cite{resh-su3,Hubbard-norms}, and it is not clear how to obtain the
form factors.

In Section \ref{sec:quench} we showed that our general result also
applies to quantum quench problems and we provided an independent
and derivation of the results of Fioretto and Mussardo
\cite{davide}. In the context of quench problems our expansion \eqref{LMsajat} can
be interpreted as a proof of a ``Generalized Eigenstate Thermalization
Hypothesis''. It would be interesting to check whether the assumption of the
single dominant root distribution holds in 
other quench situations.

In this work we only considered the one-point functions of local
operators. It is very natural to suspect that similar results hold for
multi-point correlation functions, however our methods do not apply in
that case. Concerning the non-relativistic Bose gas it can be shown along the lines of
\cite{korepin-izergin,korepin-LL1,iz-kor-resh} that equal-time two-point
functions in excited states only depend on the distribution of roots,
however there is a non-trivial dressing involved for the intermediate
state momenta. It is not clear how to obtain similar results in the
relativistic setting. In the case of finite temperature two-point
functions (with zero chemical potential) the first few terms of the
form factor expansion have been
derived recently in \cite{D22}, but it is not yet clear whether the
series can be re-summed to a compact expression.

\bigskip

\subsubsection*{Acknowledgments}

We are grateful to G\'abor Tak\'acs, M\'arton Kormos, Jean-S\'ebastien
Caux, Davide Fioretto, Guillaume Palacios, Jorn Mossel and Giuseppe Mussardo for useful
discussions. We are especially indebted to Davide Fioretto, M\'arton
Kormos and
G\'abor Tak\'acs for valuable comments about the manuscript.
The author was supported 
by the Stichting voor Fundamenteel Onderzoek der
Materie (FOM) in the Netherlands.

\newpage

{\bf\Large Appendix}

\appendix

\section{Ratios of Gaudin determinants in the thermodynamic limit}

\label{ratioss}

In this appendix we consider the thermodynamic limit of the ratios
\begin{equation*}
 \frac{\bar\rho_{N-1}(\theta_1,\dots,\hat\theta_j,\dots,\theta_N|\theta_j)}
{\rho_{N}(\theta_1,\dots,\theta_N)}
\qquad\text{and}\qquad
 \frac{\rho_{N-1}(\theta_1,\dots,\hat\theta_j,\dots,\theta_N)}
{\rho_{N}(\theta_1,\dots,\theta_N)}
\end{equation*}
The determinants above are defined in the main text.

As a first step we write the matrix $\mathcal{J}$ defined in \eqref{Jdef} as 
\begin{equation*}
  \mathcal{J}=G\Theta\qquad\text{where}
\end{equation*}
\begin{equation*}
  \Theta_{ij}=\delta_{ij} \vartheta_j,\qquad\qquad
  G_{ij}=\delta_{ij}-\frac{\varphi(\theta_{jk})}{\vartheta_j}
 \end{equation*}
\begin{equation*}
 \vartheta_j=mL\cosh\theta_j+\sum_{i=1}^N \varphi(\theta_{jk})
\end{equation*}
With this notation 
\begin{equation*}
\rho_N(\theta_1,\dots,\theta_N)=\det G_N \det \Theta_N
\end{equation*}
and also
\begin{equation*}
\rho_{N-1}(\theta_1,\dots,\hat\theta_j,\dots,\theta_N)=\det G_{N-1} \det \Theta_{N-1}
\end{equation*}
The matrix $\Theta$ is diagonal, therefore the factorization
property also applies to the sub-determinant:
\begin{equation*}
  \bar\rho_{N-1}(\theta_1,\dots,\hat\theta_j,\dots,\theta_N|\theta_j)
\det \bar G_{N-1} \det \bar \Theta_{N-1}
\end{equation*}
where $\bar G_{N-1}$ is obtained from $G_{N-1}$ simply by erasing the row and
column corresponding to $\theta_j$.

First we consider the ratio
\begin{equation*}
  \frac{\bar\rho_{N-1}\big(\theta_1,\dots,\hat
    \theta_j,\dots,\theta_N|\theta_j\big)}{\rho_N(\theta_1,\dots,\theta_N)}=
\frac{\det \bar \Theta_{N-1}}{\det  \Theta_{N}}
\frac{\det \bar G_{N-1}}{\det  G_{N}}=
\frac{1}{\vartheta_j} \frac{\det \bar G_{N-1}}{\det  G_{N}}
\end{equation*}
In the $L\to\infty$ limit we have from \eqref{Lieb}
\begin{equation*}
  \vartheta_j \to 2\pi L\rho(\theta_j)
\end{equation*}
The elements of $G_N$ can be written asymptotically as
\begin{equation*}
  G_{ij}=\delta_{ij}-\frac{1}{L}\frac{\varphi(\theta_{jk})}{\rho(\theta_j)}
\end{equation*}
The limit of $\det G_{N}$ is given by the Fredholm determinant
\begin{equation*}
  \det\Big(\hat 1-\frac{1}{2\pi}\hat K\Big)
\end{equation*}
where
\begin{equation*}
  \big(\hat K (f)\big)(x)=\int\frac{dy}{2\pi} \varphi(x-y) f(y)
\end{equation*}
Intuitively it is clear that this determinant should not change if we
erase one ``discretization point'' given by $\theta_j$. This would
then result in 
\begin{equation}
\label{detek1}
   \frac{\det\bar G_{N-1}}{\det G_{N}}=1+\ordo\left(\frac{1}{L}\right)
\end{equation}
However, we can be more precise about this. Let us decompose $\det
G_{N}$ as a single sum of sub-determinants along the column $j$:
\begin{equation*}
  \det G_{N}=\left(1-\varphi(0)\frac{1}{L \rho(\theta_j)}\right)\det \bar G_{N-1}-
\mathop{\sum_{l=1}^N}_{l\ne j}\varphi(\theta_{jl})\frac{1}{L\rho(\theta_l)} \det G_{N-1}^{(l)} 
\end{equation*}
There are $\ordo(L)$ terms in the sum on the right hand side, however
the sum itself is still of $\ordo(1/L)$ because the sub-determinants
$\det G_{N-1}^{(l)}$ are $\ordo(1/L)$ too. This proves
\eqref{detek1} and for the ratio in question we obtain
\begin{equation*}
\lim   \frac{\bar\rho_{N-1}\big(\theta_1,\dots,\hat
    \theta_j,\dots,\theta_N|\theta_j\big)}{\rho_N(\theta_1,\dots,\theta_N)}=
\frac{1}{ 2\pi L\rho(\theta_j)}
\end{equation*}

The ratio 
\begin{equation*}
  \frac{\rho_{N-1}\big(\theta_1,\dots,\hat
    \theta_j,\dots,\theta_N\big)}{\rho_N(\theta_1,\dots,\theta_N)}=
\frac{\det \Theta_{N-1}}{\det \Theta_{N}} \frac{\det G_{N-1}}{\det G_{N}}
\end{equation*}
is more involved because $\Theta_{N-1}$ and $G_{N-1}$ are not
restrictions of $\Theta_{N}$ and $G_{N}$; rather we have to subtract
all contributions associated to  $\theta_j$. 
Following the corresponding calculation in \cite{korepin-LL1} we write
\begin{equation*}
  \det \Theta_N= \vartheta_j \det \Theta_{N-1}
  \mathop{\prod_{l=1}^N}_{l\ne j} \frac{\vartheta_l}{\vartheta_l-\varphi(\theta_{jl})}
\end{equation*}
In the thermodynamic limit we have
\begin{equation*}
  \frac{\det \Theta_{N-1}}{\det \Theta_{N}}=\frac{1}{L\rho(\theta_j)}
 \mathop{\prod_{l=1}^N}_{l\ne j}
 \left(1-\frac{\varphi(\theta_{jl})}{L\rho(\theta_l)}\right)
\quad\to\quad
\frac{1}{2\pi L\rho(\theta_j)} \omega(\theta_j)
\end{equation*}
where  we  introduced the function
\begin{equation*}
  \omega(\theta)=\exp\left(-\int \frac{d\theta'}{2\pi} f(\theta')\varphi(\theta-\theta')\right)
\end{equation*}
It is not hard to convince ourselves that
\begin{equation}
\label{detek2}
   \frac{\det G_{N-1}}{\det G_{N}}=1+\ordo\left(\frac{1}{L}\right)
\end{equation}
Putting everything together
\begin{equation*}
  \lim   \frac{\rho_{N-1}\big(\theta_1,\dots,\hat
    \theta_j,\dots,\theta_N\big)}{\rho_N(\theta_1,\dots,\theta_N)}=
\frac{\omega(\theta_j)}{ 2\pi L\rho(\theta_j)}
\end{equation*}

\section{Form factors in the 1D Bose gas}

\label{sec:ABAproofs}

Here we present a detailed derivation of the singularity property
\eqref{armin}. Consider the scalar product of two arbitrary states (not necessarily
eigenvectors):
\begin{equation*}
  \bra{0}\prod_{j=1}^N \ABAC(\lambda_j^C)\prod_{k=1}^N \ABAB(\lambda_k^B)\ket{0}
\end{equation*}
The scalar product has a simple pole as $\lambda^C_N\to \lambda_N^B$:
\begin{equation}
\label{THC}
  \begin{split}
    & \bra{0}\prod_{j=1}^N \ABAC(\lambda_j^C)\prod_{k=1}^N
     \ABAB(\lambda_k^B)\ket{0}|_{\lambda^C_N\to \lambda_N^B}
\to\\& \frac{ic}{\lambda_N^C-\lambda_N^B}(l_N^C-l_N^B)
\left(\prod_{j=1}^{N-1} f_{Nj}^B f_{Nj}^C\right)\times
   \bra{0}\prod_{j=1}^{N-1} \ABAC(\lambda_j^C)\prod_{k=1}^{N-1}
     \ABAB(\lambda_k^B)\ket{0}_{\mathrm{mod}}
  \end{split}
\end{equation}
Here the scalar product on the r.h.s. has to be calculated with the
modified vacuum eigenvalues
\begin{equation*}
  a_{\mathrm{mod}}(\lambda)=a(\lambda)f(\lambda,\lambda_N)\qquad
 d_{\mathrm{mod}}(\lambda)=d(\lambda)f(\lambda_N,\lambda)
\end{equation*}
It should be noted that the remaining rapidities
$\{\lambda_1,\dots,\lambda_{N-1}\}$ satisfy the modified Bethe
equations with $l_{\mathrm{mod}}=a_{\mathrm{mod}}/d_{\mathrm{mod}}$.

Equation \eqref{THC} is valid for arbitrary vacuum eigenvalues
$a(\lambda)$ and $d(\lambda)$. In the physical case the residue
vanishes because
\begin{equation*}
  \frac{ic}{\lambda_N^C-\lambda_N^B}(l_N^C-l_N^B)\quad\to\quad
c\ l_N z_N
\end{equation*}
where we defined 
\begin{equation*}
  z(\lambda)=i\frac{\partial }{\partial \lambda}\log l(\lambda)
\end{equation*}
In the Bose gas one has $z(\lambda)=L$, however it is useful to leave
$z(\lambda)$ unspecified. 

In the diagonal case, when $\lambda_j^C\to\lambda_j^B$ for every $j$
the scalar product (the norm) depends explicitly on the variables
$l_j=l(\lambda_j)$ and $z_j=z(\lambda_j)$. The dependence on $z_N$ is
linear and one obtains from \eqref{THC}
\begin{equation}
\label{THC2}
  \frac{\partial}{\partial z_N}
 \bra{0}\prod_{j=1}^N \ABAC(\lambda_j^C)\prod_{k=1}^N
     \ABAB(\lambda_k^B)\ket{0}=
c\ l_N\left(\prod_{j=1}^{N-1} f_{Nj}^B f_{Nj}^C\right)\times
   \bra{0}\prod_{j=1}^{N-1} \ABAC(\lambda_j^C)\prod_{k=1}^{N-1}
     \ABAB(\lambda_k^B)\ket{0}_{\mathrm{mod}}
\end{equation}
Equation \eqref{THC2} was used in \cite{korepin-norms} to prove the norm formula \eqref{normalis}.

Let us consider the action of the field operator $\Psi=\Psi(0)$ on
Bethe states. It is given by \cite{iz-kor-resh}
\begin{equation}
  \Psi \prod_{k=1}^N \ABAB(\lambda_k)\ket{0}=
-i \sqrt{c} \sum_{k=1}^N l(\lambda_k) \left(\mathop{\prod_{m=1}^N}_{m\ne k}
  f(\lambda_k,\lambda_m)\right)
 \mathop{\prod_{m=1}^N}_{m\ne k} \ABAB(\lambda_m)\ket{0} 
\label{field}
\end{equation}
We also need the action of $\Psi^\dagger$:
\begin{equation}
  \bra{0} \prod_{k=1}^N \ABAC(\lambda_k) \Psi^\dagger  =
i \sqrt{c} \sum_{k=1}^N \left(\mathop{\prod_{m=1}^N}_{m\ne k}
  f(\lambda_m,\lambda_k)\right)
\bra{0} \mathop{\prod_{m=1}^N}_{m\ne k} \ABAC(\lambda_m)
\label{field2}
\end{equation}
Note that contrary to \eqref{field} the function $l(\lambda)$ is
not present in the pre-factor in \eqref{field2}. This is due to the
normalization of the operators $\ABAB(\lambda)$ and $\ABAC(\lambda)$.

We also need the multiple action of the field operators, which is
given by
\begin{equation}
\label{psi2}
  \Psi^k \prod_{j=1}^N \ABAB(\lambda_j)\ket{0}=- (m!) c
  \mathop{\sum_{\{\lambda^+\}\cup\{\lambda^-\}}}
\left(\prod_{j=1}^m l(\lambda^+_j)\right)
\left(\prod_{j=1}^m \prod_{o=1}^{N-m}
  f(\lambda^+_j,\lambda^-_o)   \right)
\mathop{\prod_{o=1}^{N-m}}  \ABAB(\lambda^-_o)\ket{0} 
\end{equation}
Here the summation runs over all bipartite partitions
$\{\lambda\}_N=\{\lambda^+\}_m\cup\{\lambda^-\}_{N-m}$ and we made use
of the identity
\begin{equation*}
  \sum_\sigma \left(\prod_{i<j} f(\lambda_{\sigma_i},\lambda_{\sigma_j})\right)=m!,
\end{equation*}
where the summation runs over all permutations $\sigma\in S_m$. The
generalization of \eqref{field2} follows straightforwardly.

The analytic properties of the (un-normalized) matrix elements
\begin{equation*}
M^k_N(\{\lambda^C\}_N,\{\lambda^B\}_N,\{l^C\}_N,\{l^B\}_N)=
\bra{0}\prod_{j=1}^N \ABAC(\lambda_j^C)(\Psi^\dagger)^k \Psi^k 
\prod_{k=1}^N \ABAB(\lambda_k^B)\ket{0}  
\end{equation*}
follow from \eqref{psi2} and the basic formula \eqref{THC}. The residue
at $\lambda^C_N\to\lambda_N^B$ is given by
\begin{equation}
\label{amstel}
  \begin{split}
    & M^k_N(\{\lambda^C\}_N,\{\lambda^B\}_N,\{l^C\}_N,\{l^B\}_N)|_{\lambda^C_N\to \lambda_N^B}
\to\\& \frac{ic}{\lambda_N^C-\lambda_N^B}(l_N^C-l_N^B)
\left(\prod_{j=1}^{N-1} f_{Nj}^B f_{Nj}^C\right)\times
M^k_{N-1}(\{\lambda^C\}_{N-1},\{\lambda^B\}_{N-1},\{l^C_{\mathrm{mod}}\}_{N-1},\{l^B_{\mathrm{mod}}\}_{N-1})
  \end{split}
\end{equation}
The modified $l$-functions are given by
\begin{equation}
\label{lmod}
  l_{\mathrm{mod}}(\lambda)=l(\lambda)\frac{f(\lambda,\lambda_N)}{f(\lambda_N,\lambda)}
\end{equation}
The form factors $\mathbb{F}_N^k$ are obtained from $M_N^k$ in the case
when two vectors are
solutions to the Bethe equations and the variables $l_j$
are expressed in terms of the functions
$f(\lambda_i,\lambda_j)$. 
The singularity properties then follows from \eqref{amstel}:
\begin{equation}
\label{armin2}
  \begin{split}
    &  \mathbb{F}^k_N(\{\lambda^C\}_N,\{\lambda^B\}_N)|_{\lambda^C_N\to \lambda_N^B}
\to\\& \frac{ic}{\lambda_N^C-\lambda_N^B}
\left(\prod_{j=1}^{N-1} f_{jN}^C f_{Nj}^B -
\prod_{j=1}^{N-1} f_{jN}^B f_{Nj}^C
\right)\times
\mathbb{F}^k_{N-1}(\{\lambda^C\}_{N-1},\{\lambda^B\}_{N-1})
  \end{split}
\end{equation}
Here we relied on the fact that the rules for the modification of
$l_j^{B,C}$ on the r.h.s. of \eqref{amstel} automatically produce the $N-1$
particle form factor. 

\section{Mean values in the 1D Bose gas}

\label{sec:mv}

Here we consider the diagonal limit 
\begin{equation*}
 M^k_{N}(\{\lambda\}_N,\{l\}_N,\{z\}_N)=
\lim_{\lambda^C_j\to \lambda_j^B}  M^k_{2N}(\{\lambda^C\}_N,\{\lambda^B\}_N,\{l^C\}_N,\{l^B\}_N),
\end{equation*}
which depends on $3N$ independent variables
$\{\lambda\}_N$, $\{l\}_N$ and $\{z\}_N$. The dependence on $z_N$ is
linear and it is given by the residue \eqref{amstel}:
\begin{equation}
\label{amstel2}
  \begin{split}
    & \frac{\partial }{\partial z_N} 
M^k_{N}(\{\lambda\}_N,\{l\}_N,\{z\}_N)=
\\& cl_N
\left(\prod_{j=1}^{N-1} f_{Nj}^B f_{Nj}^C\right)\times
M^k_{N-1}(\{\lambda\}_{N-1},\{l_{\mathrm{mod}}\}_{N-1},\{z_{\mathrm{mod}}\}_{N-1})
  \end{split}
\end{equation}
The modification rule for the variables $z_j$ follows from
\eqref{lmod} and is given by
\begin{equation}
\label{zmod}
  z_{\mathrm{mod}}(\lambda)=z(\lambda)+\varphi(\lambda-\lambda_N)
\end{equation}
The ``on-shell'' mean value
\begin{equation*}
\vev{\mathcal{\ordo}_k}_N(\{\lambda\}_N,\{z\}_N)=M^k_{N}(\{\lambda\}_N,\{l\}_N,\{z\}_N)
\end{equation*}
is a function of $2N$ parameters $\{\lambda\}_N$ and $\{z\}_N$. The dependence on $z_N$ is
linear and the coefficient is
\begin{equation}
\label{arminonly}
  \begin{split}
     \frac{\partial }{\partial z_N} 
\vev{\mathcal{\ordo}_k}_N(\{\lambda\}_N,\{z\}_N)=
 c
\left(\prod_{j=1}^{N-1} f_{jN} f_{Nj}\right)\times
\vev{\mathcal{O}_k}_{N-1}(\{\lambda\}_{N-1},\{z_{\mathrm{mod}}\}_{N-1})
  \end{split}
\end{equation}
The rule for the modification of the variables $z_j$ is given by
\eqref{zmod}.

It is convenient to define the symmetric evaluation of the diagonal form factor
as
\begin{equation*}
   \mathbb{F}^k_{2N,s}(\{\lambda\}_N)=\lim_{\eps\to 0}
 \mathbb{F}^k_{2N}(\{\lambda^B+\eps\}_N,\{\lambda^B\}_N)
\end{equation*}
The function $\mathbb{F}^k_{2N,s}$ is the ``irreducible part'' of the
diagonal matrix element; it differs from the $F_{N,s}^k$ defined in
the main text only in the over-all normalization.

\begin{thm}
\label{thm-ss}
The irreducible part is equal to the mean value at $z_j=0$:
  \begin{equation}
 \mathbb{F}^k_{2N,s}(\{\lambda\}_N)=\vev{\mathcal{O}_k}_N(\{\lambda\}_N,\{0\}_N)
  \end{equation} 
\end{thm}

{\bf Proof} It is clear from \eqref{amstel} that the $z_N$ dependence
of the mean value arises from the rapidity dependence of the function
$l(\lambda)$. Therefore, the irreducible part can be obtained by
putting $l^C_j=l^B_j$ where $\{l^B\}$ solves the Bethe equations:
\begin{equation*}
 \vev{\mathcal{O}_k}_N(\{\lambda\}_N,\{0\}_N)=\lim_{\eps\to 0}
 M^k_{2N}(\{\lambda^B+\eps\}_N,\{\lambda^B\}_N,\{l^B\}_N,\{l^B\}_N)
\end{equation*}
On the other hand, the form factor is obtained by
\begin{equation}
\label{keki}
   \mathbb{F}^k_{2N,s}(\{\lambda\}_N)=
\lim_{\eps\to 0} M^k_{2N}(\{\lambda^B+\eps\}_N,\{\lambda^B\}_N,\{\tilde l^B\}_N,\{l^B\}_N),
\end{equation}
where the Bethe equations are satisfied. This means that $\{\tilde
l^B\}_N$ and $\{l^B\}_N$ will be given by the appropriate products of
$S$-matrices. However, a constant shift in the rapidities yields 
\begin{equation*}
  \{\tilde l^B\}_N=\{l^B\}_N
\end{equation*}
This proves the theorem.
\bizveg

\bigskip

For any partition $\{\lambda\}_N=\{\lambda^+\}_{n}\cup \{\lambda^-\}_{N-n}$ we define the function
\begin{equation*}
\begin{split}
&  S_{N}(\{\lambda^+\}_{n},\{\lambda^-\}_{N-n},\{z\}_N)=\\
&\hspace{1cm}c^N \left(\prod_{j=1}^n \prod_{k=1}^{N-n}
 |f(\lambda_j^+-\lambda_k^-)|^2
\right)
\left(\prod_{1\le j < k \le n} |f(\lambda_j^--\lambda_k^-)|^2
\right) 
\rho_{N-n}(\{\lambda^-\}_{N-n},\{z\}_N)
\end{split}
\end{equation*}
The dependence on $z_N$ is given by
\begin{equation}
\label{scooter}
\begin{split}
&     \frac{\partial }{\partial z_N} 
 S_{N}(\{\lambda^+\}_{n},\{\lambda^-\}_{N-n},\{z\}_N)=\\
&\hspace{1cm}= \begin{cases}
    c
\left(\prod_{j=1}^{N-1} f_{jN} f_{Nj}\right)\times
 S_{N-n}(\{\lambda^-\}_{N-n},\{z_{mod}\}_N) & \text{if } \lambda_N\in
 \{\lambda^-\}_{N-n}\\
0 & \text{if }  \lambda_N\in
 \{\lambda^+\}_{n}
 \end{cases}
\end{split}
\end{equation}

Now we are in a position to prove the main theorem of this appendix:

\bigskip

\begin{thm}
\label{meanvaluetheorem}
  The mean value can be expressed as
\begin{equation}
\label{meanvalue}
   \vev{\ordo_k}_N=\sum_{\{\lambda^+\}\cup\{\lambda^-\}}
   \mathbb{F}^k_{n,s}(\{\lambda^+\})\ S_{N-n}(\{\lambda^-\})
\end{equation}
where the summation is over the bipartite partitions of the rapidities
into two subsets $\{\lambda\}_N=\{\lambda^+\}_n\cup \{\lambda^-\}_{N-n}$.
\end{thm}
{\bf Proof} The proof is given by induction over $N$. The form factor vanishes
for $N<k$, therefore the first member will be at $N=k$. In this case
there is no dependence on any of the $z_j$, therefore \eqref{meanvalue} is
satisfied. Now let us assume that \eqref{meanvalue} is proven for
$N<M$. We prove that it is valid for $N=M$.

We investigate the $z_j$ dependence of  \eqref{meanvalue}.
According
to the assumption of induction it follows from \eqref{arminonly} and
\eqref{scooter} that the $z_j$ dependence of the l.h.s. and r.h.s. in
\eqref{meanvalue} coincides. Therefore we consider the value
at $z_j=0$. At this point the norm functions the r.h.s. of
\eqref{meanvalue} vanish and one is left
with 
\begin{equation}
   \mathbb{F}^k_{M,s}(\{\lambda\})
\end{equation}
It was proven in Theorem 1 that this form factor coincides with the
irreducible part of the mean value. Thus \eqref{meanvalue} is
proven. \bizveg

The normalized expression of \eqref{fftcsa2-result-LL} follows from
\eqref{meanvalue} by
dividing with the norm \eqref{normalis}.

\section{The form factors $F^1_{2N,s}$ and  $F^2_{2N,s}$}

\label{sec:FFs}

In this section we derive explicit expressions for the functions
$F_{2N,s}^k$ and $F_{2N,c}^k$ for $N\in\mathbb{N}$ and $k=1,2$. 

In the
first case the operator in question is $\ordo_1=\Psi^\dagger\Psi$
which describes the total particle density. Therefore its mean value is
given by
\begin{equation}
\label{g1a}
  \vev{\Psi^\dagger \Psi}_N=\frac{N}{L}=\frac{1}{\rho_N(\{\lambda\})}
  \frac{N \rho_N(\{\lambda\})}{L}
\end{equation}
On the other hand Theorem \ref{meanvaluetheorem} gives
\begin{equation}
\label{g1b}
\begin{split}
    \vev{\Psi^\dagger \Psi}_N=&\frac{1}{\rho_N(\{\lambda\})}
\sum_{\{\lambda_+\}\cup
      \{\lambda_-\} }
F^1_{2k,s}(\{\lambda_-\})
\rho_{N-k}(\{\lambda_+\})
\end{split}
\end{equation}
Comparing \eqref{g1a} and \eqref{g1b} one can inductively deduce the
form factors $F_{2N}^1$. The simplest way is to apply Theorem
\ref{thm-ss} which yields
\begin{equation*}
 F^1_{2N,s}(\{\lambda\})=\left. \frac{N \rho_N(\{\lambda\})}{L}\right|_{L=0}
\end{equation*}
Let us introduce the matrix $J$ which is given by $\rho_{N}$ at
$L=0$:
\begin{equation*}
  J_{ij}=\delta_{ij} \big(\sum_{k=1}^n \varphi_{ik}\big)-\varphi_{ij}
\end{equation*}
It is easy to see that 
\begin{equation*}
 \left.  \frac{\rho_{N}\big(\{\lambda\}\big)}{L}\right|_{L=0}=
\sum_{i=1}^N \bar J_{ii}
\end{equation*}
where the elements of $\bar J$ are the corresponding minors of
$J$. The matrix $J$ has the property that the sums of its rows and the
sums of its columns give zero, therefore its minors
coincide. Therefore
\begin{equation*}
  F_{2N,s}^1\big(\{\lambda\}_N\big)=N^2 \bar J_{11}
\end{equation*}
In graph theory the matrix $J$ is the generalized Laplacian matrix of $G_n$
 (the complete graph with $n$ edges), where for each vertex from node $i$
to $j$ one associates the ``weight'' $\varphi_{ij}$. Then according to
Kirchhoff's theorem the minor $\bar J_{11}$ gives the enumeration of
the spanning trees of $G_n$:
\begin{equation*}
  \bar J_{11}= \mathcal{G}_N(\theta_1,\dots,\theta_N),
\end{equation*}
where
\begin{equation*}
\mathcal{G}_N(\theta_1,\dots,\theta_N) =
\sum_{G\in ST} \left(\prod_{\alpha\in  V_G} \varphi(\alpha)\right)
\end{equation*}
The sum runs over all spanning trees and the product runs over
the vertices $\alpha$ of a given spanning tree, and 
\begin{equation*}
  \varphi(\alpha)=\varphi_{ij}
\end{equation*}
for a vertex $\alpha$ going from edge $i$ to $j$. Putting everything
together
\begin{equation*}
  F_{2N,s}^1\big(\{\lambda\}_N\big)=N^2 \sum_{G\in ST} \left(\prod_{\alpha\in  V_G} \varphi(\alpha)\right)
\end{equation*}
By the reverse application of Theorem \ref{par:Theorem-1} one can deduce that
\begin{equation}
\label{llg1}
  F^{1}_{2N,c}(\lambda_1,\dots,\lambda_N)=\sum_\sigma \varphi(\lambda_{\sigma_1}-\lambda_{\sigma_2})
  \varphi(\lambda_{\sigma_2}-\lambda_{\sigma_3}) \dots
\varphi(\lambda_{\sigma_{N-1}}-\lambda_{\sigma_N}) 
\end{equation}
Here the summation runs over all permutations $\sigma\in S_N$. 

In the case of the operator $\ordo_2=\Psi^\dagger\Psi^\dagger\Psi
\Psi$ one makes use of the fact that it is just the
interaction term in the Hamiltonian. Then the Hellmann-Feynman theorem
gives \cite{Hellmann-Feynman}
\begin{equation*}
    \vev{\Psi^\dagger\Psi^\dagger\Psi\Psi}_N=\frac{1}{L}
\frac{\partial E(L,c)}{\partial c}
\end{equation*}
Here $E(L,c)$ is the total energy of the $N$-particle state for 
fixed momentum quantum numbers. 
It follows from scaling arguments that 
\begin{equation*}
  E(L,c)=\frac{h(Lc)}{L^2},
\end{equation*}
where $h(x)$ is a dimensionless function. Therefore
\begin{equation*}
  \frac{1}{L}\frac{\partial E}{\partial c}=\frac{1}{c}
\left(\frac{\partial E}{\partial L}+2\frac{E}{L}\right)
\end{equation*}
The energy is given by
\begin{equation*}
  E=\sum_i \lambda_i^2\qquad \text{and}\qquad
\frac{\partial E}{\partial L}=2\lambda_i \frac{\partial \lambda_i}{\partial L}
\end{equation*}
The derivative with respect to $L$ can be calculated from the
Bethe-equations:
\begin{equation*}
\lambda_i+\mathcal{J}_{ij} \frac{\partial \lambda_j}{\partial L}=0
\end{equation*}
Therefore
\begin{equation*}
    \vev{\Psi^\dagger\Psi^\dagger\Psi\Psi}_N=
\frac{1}{\rho_N}\frac{2}{c}\left[
\lambda_i \left(-\bar{\mathcal{J}}_{ij}+\delta_{ij}\frac{\rho_N}{L}\right)\lambda_j
\right]
\end{equation*}
Making use of Theorem \ref{thm-ss} one arrives at
\begin{equation*}
  F_{2N,s}^2(\theta_1,\dots,\theta_N)=\frac{2}{c}\mathcal{G}_N(\theta_1,\dots,\theta_N) 
\times\left(\sum_{i<j} (\lambda_i-\lambda_j)^2\right)
\end{equation*}
Theorem \ref{par:Theorem-1} then yields
\begin{equation}
\label{llg2}
  F^{2}_{2N,c}(\lambda_1,\dots,\lambda_N)=\frac{1}{c}\sum_\sigma \varphi(\lambda_{\sigma_1}-\lambda_{\sigma_2})
  \varphi(\lambda_{\sigma_2}-\lambda_{\sigma_3}) \dots
\varphi(\lambda_{\sigma_{N-1}}-\lambda_{\sigma_N}) (\lambda_{\sigma_1}-\lambda_{\sigma_N})^2
\end{equation}
The summation runs over all permutations $\sigma\in S_N$. The results
\eqref{llg1} and \eqref{llg2} are in agreement with the formulas of
\cite{sinhG-LL2} presented for $N=1,2,3$.

\section{The $x\to 0$ limit of the non-local correlation function}

\label{sec:xto0}
In this appendix we demonstrate how to re-derive the expectation
values $\vev{\ordo_k}$
from the previously
available results for non-local correlation 
functions \cite{korepin-LL1}. We only consider
$\ordo_2={\Psi^\dagger\Psi^\dagger\Psi\Psi}$, which is related to the 
operator $Q^2(x)$.
The normal ordering of the field operators yields:
\begin{equation*}
  Q^2(x)-Q(x)=\int_0^x\int_0^x dx_1 dx_2\  \Psi^\dagger(x_1)\Psi^\dagger(x_2)\Psi(x_1)\Psi(x_2)
\end{equation*}
Taking the limit $x\to 0$ one gets:
\begin{equation}
\label{eakarjuk}
   \Psi^\dagger\Psi^\dagger\Psi\Psi=\lim_{x\to 0} \frac{
     Q^2(x)-Q(x)}{x^2}
\end{equation}
Therefore, in order to obtain the expectation value of $\ordo_2=
\Psi^\dagger\Psi^\dagger\Psi\Psi$ one has to pick the $\ordo(x^2)$ term
from the expansion of $Q^2(x)-Q(x)$. We perform this for the
expectation values with a finite number of particles. 
The relevant result reads
(eq. 3.32 from \cite{korepin-LL1})
\begin{equation}
\label{Q2a}
  \vev{(Q(x))^2}_N=\frac{1}{\rho_N(\{\lambda)\}} \left[
\vev{(Q(x))^2}_N^0+\sum_{\{\lambda_+\}\cup
      \{\lambda_-\} }
  I_{n,N}(\{ \lambda_- \}_n,\{\lambda_{+}\}_{N-n})
\rho_{N-n}(\{\lambda_+\})\right]
\end{equation}
Here the summation runs over the partitions with $|\{\lambda_-\}|=n$
with $n=2\dots N$. The first term is given by the ``irreducible part
of the identity operator''  (eq. 3.28 in \cite{korepin-LL1})
\begin{equation}
\label{Q2i}
  \vev{(Q(x))^2}_N^0=\sum_{\{\lambda_x\}\cup\{\lambda_y\} }
n^2\  \rho_{n}^x(\{\lambda_x\})  \rho_{N-n}^y(\{\lambda_y\})
\end{equation}
Here $\rho_{n}^x(\{\lambda_x\})$ and  $\rho_{N-n}^y(\{\lambda_y\})$
are the Gaudin-determinants of a system with volume parameters $x$ and $y=L-x$,
respectively. The quantities $ I_{n,N}(\{ \lambda_-
\}_n,\{\lambda_{+}\}_{N-n})$ are the ``irreducible parts'' of the
operator $Q^2$. They have a non-trivial dependence on $x$. In the case
of $Q(x)$ the ``irreducible parts'' are zero and the 
mean value can be expressed as (eq. 3.23 in \cite{korepin-LL1})
\begin{equation}
\label{g1aa}
  \vev{Q(x)}_N=\frac{1}{\rho_N(\{\lambda)\}}\sum_{\{\lambda_x\}\cup\{\lambda_y\} }
n\  \rho_{n}^x(\{\lambda_x\})  \rho_{N-n}^y(\{\lambda_y\})
\end{equation}
In the following we only consider the first non-trivial case $N=2$. 
Equations \eqref{Q2a},
\eqref{Q2i} and \eqref{g1aa} yield
\begin{equation*}
      \vev{(Q(x))^2-Q(x)}_2=
\frac{1}{\rho_2(\lambda_1,\lambda_2)}\Big(
2\rho_2^x(\lambda_1,\lambda_2)+I_2(\lambda_1,\lambda_2)
\Big)
\end{equation*}
Here
\begin{equation*}
  \rho_2^x(\lambda_1,\lambda_2)=x(x+2\varphi_{12})
\end{equation*}
and $I_2$ is given  by 8.11-8.12 in \cite{korepin-izergin}:
\begin{equation*}
  I_2=-\frac{2}{\lambda_{12}^2}\left[
(e^{-ix\lambda_{12}}-1)\frac{\lambda_{12}+ic}{\lambda_{12}-ic}+
(e^{ix\lambda_{12}}-1)\frac{\lambda_{12}-ic}{\lambda_{12}+ic}
\right]
\end{equation*}
Expanding in $x$ and keeping only the first two terms one has
\begin{equation*}
  I_2  =-4x\varphi_{12}+x^2\Big(\frac{2\varphi_{12}\lambda_{12}^2}{c}-2\Big)+\dots
\end{equation*}
Putting everything together one observes that the $\ordo(x)$ terms
indeed cancel 
and one is left with
\begin{equation*}
     \vev{(Q(x))^2-Q(x)}_2=
\frac{x^2}{\rho_2(\lambda_1,\lambda_2)}
\frac{2\varphi_{12}\lambda_{12}^2}{c}
\end{equation*}
Therefore
\begin{equation*}
    \vev{\Psi^\dagger\Psi^\dagger \Psi\Psi}_2=\frac{1}{\rho_2(\lambda_1,\lambda_2)}
\frac{2\varphi_{12}\lambda_{12}^2}{c}.
\end{equation*}
Comparing with \eqref{fftcsa2-result-LL} we conclude that
\begin{equation*}
F^2_{2,s}=0,\qquad\quad
  F^2_{4,s}=\frac{2\varphi_{12}\lambda_{12}^2}{c}
\end{equation*}
The above formulas are in complete agreement with our result
\eqref{llg2}.

\addcontentsline{toc}{section}{References}
\bibliography{../pozsi-general}
\bibliographystyle{utphys}

\end{document}